\documentclass[letterpaper,11pt]{article}

\usepackage[utf8]{inputenc}

\usepackage[top=3.25cm, bottom=3cm, left=3.5cm, right=3.5cm,           heightrounded,marginparwidth=1.5cm, marginparsep=1cm]{geometry} %for margins

\usepackage[shortlabels]{enumitem} % for lists

\usepackage[square,numbers,merge,comma,sort&compress]{natbib} % bibliography
\makeatletter
\def\NAT@spacechar{\,}  % define space inside [1,\,2]
\makeatother

\usepackage{amsmath,amssymb,amsfonts,amsthm,amsbsy}
\usepackage{mathtools} % for {rcases}

\usepackage{fnpct}% for multiple footnote at same point
\setfnpct{dont-mess-around} % no other kerning and punctuation switching

\usepackage{slashed,cancel,latexsym}
\usepackage{old-arrows,comment,relsize,setspace,moresize,accents}
\usepackage{changepage,etoolbox}
\usepackage{epsfig}
\usepackage{mathrsfs,calligra,aurical,bold-extra} % fonts, \mathscr ...
\usepackage{calc,float,appendix}
\usepackage{bigints,xargs,extarrows}
\usepackage{empheq,soul} % for \hl{...}
\usepackage{upgreek}
\usepackage[all]{xy}
\usepackage[normalem]{ulem}

\usepackage[svgnames]{xcolor}  % Load BEFORE tikz!
\definecolor{Blue}{rgb}{0.0, 0.0, 0.37}
\definecolor{Green}{rgb}{0.05, 0.45, 0.25}
\xdefinecolor{dogwoodrose}{rgb}{0.8, 0.1, 0.55}
\xdefinecolor{RRed}{rgb}{0.7, 0.1, 0.4}

\usepackage{tikz}
\usetikzlibrary{scopes,decorations.pathmorphing,patterns,calc,arrows,shapes.geometric,shapes.arrows,decorations.markings,plotmarks}
\usepackage{pgfplots}

\usepackage[blocks]{authblk}  % authors and affiliations

\usepackage{graphicx} % Required for including images
\graphicspath{{Figures/}} % Set the default folder for images

\usepackage[labelsep=colon]{caption}  % captions
\usepackage[labelsep=colon,aboveskip=10pt,belowskip=10pt]{subcaption}
\DeclareCaptionSubType * [roman]{table}
\usepackage{titlesec}
\titleformat{\section}{\normalfont\fontsize{12.5}{12}\bfseries}{\thesection}{0.5em}{}%{\phantomsection}
\titleformat{\subsection}{\normalfont\fontsize{10.5}{10}\bfseries}{\thesubsection}{0.5em}{}%{\phantomsection}
\titleformat{\subsubsection}{\normalfont\normalsize\bfseries}{\thesubsubsection}{0.5em}{}%{\phantomsection}

\titlespacing*{\section}{0pt}%
                {4ex plus 1ex minus .5ex}{1.75ex plus .25ex minus .25ex}
\titlespacing*{\subsection}{0pt}%
                {3.5ex plus 1ex minus .5ex}{1.25ex plus .2ex minus .2ex}
\titlespacing*{\subsubsection}{0pt}%
                {2.5ex plus 0.75ex minus .2ex}{0.75ex plus .15ex minus .15ex}
\titlespacing*{\paragraph}{0pt}%
                {1.85ex plus 0.5ex minus .15ex}{1em}

\usepackage{titletoc}
\addtocontents{toc}{\addvspace{-0.75em}}  % reduce vert space before TOC
\usepackage{bm}  % \mathbbm only has digits "1" and "2"
\usepackage{dsfont}  % double math characters (but only digit "1")

\DeclareMathAlphabet{\mathpzc}{OT1}{pzc}{m}{it}
\DeclareMathAlphabet{\mathcal}{OMS}{cmsy}{m}{n}
\DeclareSymbolFontAlphabet{\Scr}{rsfs}
\DeclareMathAlphabet{\mathbold}{U}{BOONDOX-ds}{m}{n}
\SetMathAlphabet{\mathbold}{bold}{U}{BOONDOX-ds}{b}{n}
\DeclareMathAlphabet{\mathcalboondox}{U}{BOONDOX-calo}{m}{n}
\SetMathAlphabet{\mathcalboondox}{bold}{U}{BOONDOX-calo}{b}{n}
\DeclareMathAlphabet{\mathbcalboondox}{U}{BOONDOX-calo}{b}{n}

\DeclareFontFamily{U}{matha}{\hyphenchar\font45}
\DeclareFontShape{U}{matha}{m}{n}{ <5> <6> <7> <8> <9> <10> gen * matha
                    <10.95> matha10 <12> <14.4> <17.28> <20.74> <24.88> matha12}{}
\DeclareSymbolFont{matha}{U}{matha}{m}{n}
\DeclareMathSymbol{\varleftarrow}{3}{matha}{"D0}
\DeclareMathSymbol{\varrightarrow}{3}{matha}{"D1}
\DeclareMathSymbol{\simeq}{3}{matha}{"14}
\DeclareMathSymbol{\sim}{3}{matha}{"12}
\DeclareMathSymbol{\ll}{3}{matha}{"21}
\DeclareMathSymbol{\gtrsim}{3}{matha}{"C1}
\DeclareMathSymbol{\lesssim}{3}{matha}{"C0}

\newcommand\linkcol{RRed}

\usepackage[breaklinks=true,backref=page]{hyperref}
\hypersetup{
    bookmarks=true,         % show bookmarks bar?
    bookmarksnumbered=true,
    pdftoolbar=true,        % show Acrobat’s toolbar?
    pdfmenubar=true,        % show Acrobat’s menu?
    pdffitwindow=false,     % window fit to page when opened
    pdfauthor={},     % author
    pdfsubject={},   % subject of the document
    pdfcreator={},   % creator of the document
    pdfproducer={},  % producer of the document
    pdfpagemode={UseNone},
    pdfstartview={FitH},
    colorlinks=true,
    plainpages,
    linktoc=page,
    citecolor=blue,
    filecolor=black,
    linkcolor=\linkcol,
    urlcolor=Green,
}
\renewcommand*{\backref}[1]{}
\renewcommand*{\backrefalt}[4]{%
\ifcase #1 %
\relax
\or
~{\small [\textsc{p.~\fns{\!#2}}]}
\else
~{\small [\textsc{p.~\fns{\!#2}}]}%
\fi}

\usepackage{footnotebackref}
\usepackage{hypernat} % Load it HERE, after NATBIB and HYPERREF
\usepackage{cleveref} % Load it HERE, after HYPERREF

\makeatletter
\normalsize
\makeatother

\makeatletter
\g@addto@macro\bfseries{\boldmath}
\makeatother

\def\+{~+~}
\def\-{~-~}
\def\={\:=\:}
\newcommand\fns{\footnotesize}

\newcommandx\Hodge[1][1=4,usedefault]{{}^{\star_{#1}}}

\newcommandx{\overbar}[1]{\mkern
1.5mu\overline{\mkern-2.0mu#1\mkern-2.0mu}\mkern 1.5mu}
\newcommandx{\overbarcal}[1]{\mkern                   6.0mu\overline{\mkern-5.5mu#1\mkern-1.0mu}\mkern 1.5mu}

\newcommand{\xs}{x_{\textsc{s}}}
\newcommand{\chiR}{\chi_{{}_{\text{R}}}}
\newcommand{\chiI}{\chi_{{}_{\text{I}}}}
\newcommand{\N}{\mathcal{N}}

\newcommandx{\eM}[1][1=A,usedefault]{\epsilon_{{}_{(\text{M})}}^{#1}}

\title{%
\vspace{-1.0cm}\centering\LARGE\bfseries
       Supersymmetric solitons in gauged $\mathcal{N} = 8$ supergravity
\bigskip\bigskip}

\author[1]{Andr\'{e}s Anabal\'{o}n\footnote{\href{anabalo@gmail.com}{anabalo@gmail.com}}}

\author[2]{Antonio Gallerati\footnote{\href{mailto:antonio.gallerati@polito.it}{antonio.gallerati@polito.it}}}

\author[3]{Simon Ross\footnote{\href{simonfross@gmail.com}{simonfross@gmail.com}}}

\author[2,4]{Mario Trigiante\footnote{\href{mario.trigiante@polito.it}{mario.trigiante@polito.it}}}

\affil[1]{%\bigskip
{Departamento de Ciencias, Facultad de Artes Liberales, Universidad
Adolfo Ib\'{a}\~{n}ez, Avenida Diagonal las Torres 2640, Pe\~{n}alolen, Santiago, Chile}
\medskip
}

\affil[2]{%\bigskip
{Politecnico di Torino, Dipartimento di Scienza Applicata e Tecnologia, corso Duca degli Abruzzi 24, 10129 Torino, Italy}%
\medskip
}

\affil[3]{%\bigskip
{Centre for Particle Theory, Department of Mathematical Sciences,
Durham University, South Road, Durham DH1 3LE, U.K.}%
\medskip
}

\affil[4]{%\bigskip
{Istituto Nazionale di Fisica Nucleare, Sezione di Torino, via Pietro Giuria 1, 10125 Torino, Italy}%
\medskip
}

\date{}

\begin{document}

\maketitle
%\smallskip
%\vspace{-1.5em}

\begin{abstract}
{\noindent%
We consider soliton solutions in AdS$_{4}$ with a flat slicing and Wilson loops around one cycle. We study the phase structure and find the ground state and identify supersymmetric solutions as a function of the Wilson loops. We work in the context of a scalar field truncation of gauged $\mathcal{N}=8$ supergravity, where all the dilatons are equal and all the axions vanish in the STU model. In this theory, we construct new soliton solutions parameterized by two Wilson lines. We find that there is a degeneracy of supersymmetric solutions. We also show that, for alternate boundary conditions, there exists a non-supersymmetric soliton solution with energy lower than the supersymmetric one.}
\end{abstract}

%\bigskip
\newpage

\tableofcontents
%\pagebreak
\newpage

%-- ARTICLE SECTIONS ----

\section{Introduction and discussion}
An interesting aspect of the AdS/CFT correspondence is the study of the possible ground states as a function of the boundary conditions. The first non-trivial example is the AdS soliton of \cite{Horowitz:1998ha}, which is conjectured to correspond to the ground state of the theory with a planar boundary, with one direction compactified on a circle with antiperiodic boundary conditions for fermions. In the present paper, we study generalizations of the AdS soliton when we add Wilson loops for gauge fields around the circle.\par
We will work in the context of asymptotically AdS$_4$ solutions of a truncation of $\mathcal N=8$ supergravity. Let us consider a representative of the conformal
boundary of a four-dimensional metric as
\begin{equation}
ds_\text{bound.}^{2}=-dt^{2}+d\varphi^{2}+dz^{2}\:,%
\end{equation}
where $\varphi$ is a periodic coordinate. Fermions can be either periodic or anti-periodic around the circle parametrized by
$\varphi$. If fermions are anti-periodic, it is possible to construct an interior solution where the
vector $\partial_{\varphi}$ has vanishing norm: this is the AdS soliton. One
might wonder what happens if extra sources are added to this configuration, like, for instance, a Wilson loop
\begin{equation}
\Phi_{\textsc{m}}=\oint A_{\varphi}\:d\varphi\:.\label{WL}%
\end{equation}
In the following, we will consider the case of anti-periodic boundary conditions on the $\varphi$ circle, but periodic boundary conditions on the $z$ circle. As we will see, this allows supersymmetric solitons for appropriate choices of Wilson loops; it also has the effect of excluding the possibility of an AdS soliton solution where the $z$ circle contracts smoothly in the interior. The relevant solutions will then either have both circles non-contractible in the interior, or the $\varphi$ circle contractible. The latter solutions, which generalise the AdS soliton and can be obtained by double analytic continuation from electrically charged black holes \cite{Witten:1981gj,Horowitz:1998ha,Constable:1999gb,Balasubramanian:2002am,Birmingham:2002st,Astefanesei:2005eq,Oliva:2009ip,Stotyn:2011tv,Dibitetto:2020csn,Bah:2020ogh,Bobev:2020pjk,Anabalon:2021tua}, are the main focus of our interest. \par
In the case
of the simplest Einstein-Maxwell theory, the configuration has been known for
a while \cite{Astorino:2012zm, Kastor:2020wsm}. In \cite{Anabalon:2021tua} it was found that some of these configurations are
supersymmetric and that, at the supersymmetric point, there are two
possible solutions, the soliton and a Poincar\'e-AdS solution dressed with a constant Wilson loop. In this article, we extend this study to the case of
gauged $\mathcal{N}=8$ supergravity, and construct solutions of its STU model truncation in which the three dilatons are equal and all axions vanish. In particular, we want to analyse whether the degeneracy of supersymmetric solutions extends to this more general context.\par
In the pure Einstein-Maxwell case, supersymmetry was obtained for a single value of the source \eqref{WL} \cite{Anabalon:2021tua}. In the model we study here, there are two Wilson lines, with a one-parameter family of values of the Wilson lines which give supersymmetric solitons. There are also Poincar\'e-AdS solutions and domain wall solutions dressed by constant Wilson loops, which satisfy the same boundary conditions.\par
We observe that for the Poincar\'e-AdS solutions regularity requires a quantization condition on the Wilson loops which was missed in \cite{Anabalon:2021tua}. This arises from considering the uplift of the configuration to a $M_{4}\times S^{7}$ solution of M theory, where the Wilson loops parametrise a shift on the $S^7$ as we go around the $\varphi$ circle.

For fixed flux boundary conditions, the coexistence of the soliton and the Poincar\'e-AdS and domain wall solutions leads to a degeneracy of supersymmetric solutions as in \cite{Anabalon:2021tua}. We also  consider the alternate boundary condition of fixed currents on the boundary --  we will in the sequel refer to these boundary conditions as fixed charge -- although this is actually correct only for the Euclidean continuation, as the currents are spacelike in the Lorentzian solution.  We find that, for supersymmetry-preserving fixed charge boundary conditions, there are two distinct soliton solutions, leading to a new kind of degeneracy of supersymmetric solutions (the Poincar\'e-AdS and domain wall solutions do not satisfy these boundary conditions, so there is no degeneracy at fixed charge in the previous case studied in \cite{Anabalon:2021tua}).\par
Finally, we find that the non-supersymmetric solutions of \cite{Anabalon:2021tua} also satisfy the  boundary conditions which give supersymmetric solutions for fixed charge, and one branch of those solutions has lower energy than the supersymmetric solutions. This is surprising as we would expect the supersymmetric solutions to saturate a BPS bound, forbidding the existence of solutions with lower energy. The BPS bound for these alternate boundary conditions has however not been explicitly worked out as far as we are aware. We discuss this result in light of the \emph{positive energy theorem} \cite{Witten:1981mf,Gibbons:1983aq}, which implies that the energy of a supersymmetry preserving solution is lower than the energy of any other solution satisfying the same boundary conditions. We show that our result is not in contradiction with this general property. The central point of the argument is that a necessary condition for the positive energy theorem to apply is the existence, for the non-supersymmetric solution, of an asymptotic Killing spinor which coincides, up to $O(1/r^2)$ terms at radial infinity, with the Killing spinor of the supersymmetric one. Since the latter has antiperiodic boundary conditions along the circle at infinity, in order for the positive energy theorem to apply, the non-supersymmetric solutions should admit an asymptotic Killing spinor with the same property at the boundary. As we shall prove, this is the case only if the charges at infinity have specific values, for which the energy of the non-supersymmetric solution exceeds that of the supersymmetric one. In summary, there is no contradiction with the positive energy theorem if we include among the boundary conditions those applying to the asymptotic Killing spinor.\par
Another important direction for future work is to find a more general understanding of the degeneracy of the susy solutions. We believe this is a generic feature of such boundary conditions; we intend to provide a general proof in a forthcoming paper.\par
It is also possible to construct black holes in this theory \cite{Cacciatori:2009iz, Hristov:2012bd, Anabalon:2020pez,Anabalon:2021smx,Anabalon:2017yhv,Anabalon:2020qux,Gallerati:2021cty} and endow them with Wilson lines along the lines of \cite{Anabalon:2022ksf}. This will provide an even more complete phase diagram of this model that we leave to analyze in the future.
\par\smallskip
The outline of the paper is as follows. In the next Section \ref{sec:model}, we review the model under consideration. In Section \ref{sol}, we present the soliton solutions within this model, and explain their relation to the solutions of \cite{Anabalon:2021tua}. In Section \ref{domw}, we discuss the solutions with constant fluxes, and explain the quantisation of the flux from demanding a well-behaved action on the $S^7$ factor in the uplift. In Section \ref{susy}, we find the supersymmetric solutions for both fixed flux and fixed charge boundary conditions. In Section \ref{phase}, we describe the phase structure for the different boundary conditions, and point out that for supersymmetric fixed charge boundary conditions there are both supersymmetric and non-supersymmetric solutions, with a surprising family of non-supersymmetric solutions of lower energy and free energy than the supersymmetric ones. The consistency of this result with the positive energy theorem for asymptotically AdS solutions is discussed in subsection \ref{PET}.

\section{The model}\label{sec:model}
We are interested in studying the dilatonic sector of the STU\ model of the
$\mathrm{SO}(8)$-gauged, $\mathcal{N}=8$ supergravity with action:
\begin{equation}
\mathcal{S}=\frac{1}{2\,\kappa}\int\! d^{4}x\:\sqrt{-g}\left(R-\sum_{i=1}^{3} \frac{\left(\partial\Phi_{i}\right)  ^{2}}{2}+\frac{2}{L^{2}}\,\cosh\left(  \Phi_{i}\right)
-\frac{1}{4}\:\sum_{i=1}^{4}X_{i}^{-2}\bar{F}_{i}^{2}\right)\:,
\label{LSTU}
\end{equation}
where $\bar{F}_{i}$ are two forms, related with gauge fields in the standard
way
\begin{align}
\bar{F}_{i}=d\bar{A}_{i}\,,
\qquad
X_{i}=e^{-\frac{1}{2}\vec{a}_{i}\cdot
\vec{\Phi}}\,,
\qquad
\vec{\Phi}=\left(\Phi_{1},\Phi_{2},\Phi_{3}\right)\,,
\end{align}
and
\begin{equation}
\vec{a}_{1}=\left(1,1,1\right),
\qquad
\vec{a}_{2}=\left(1,-1,-1\right),
\qquad
\vec{a}_{3}=\left(-1,1,-1\right),
\qquad
\vec{a}_{4}=\left(-1,-1,1\right).\;
\end{equation}
We will be interested in purely magnetic solutions, in which case it is
consistent to truncate the axions to zero. The Lagrangian \eqref{LSTU} can be
obtained from the compactification of eleven dimensional supergravity over the
seven sphere with the ansatz \cite{Cvetic:1999xp}
\begin{align}
ds_{11}^{2}  =&\;\tilde{\Delta}^{2/3}\,ds_{4}^{2}+4\,L^{2}\tilde{\Delta}^{-1/3}\:\sum_{i=1}^{4}X_{i}^{-1}\left(d\mu_{i}^{2}+\mu_{i}^{2}\left(d\varphi_{i}+\frac{1}{2L}\bar{A}_{i}\right)^{2}\right)\:,
\\[1ex]
F =&-\frac{1}{L}\:\epsilon_{4}\:\sum_{i=1}^{4}\left(X_{i}^{2}\mu_{i}^{2}-\tilde{\Delta}\,X_{i}\right)  +L\,X_{i}^{-1}\:\Hodge dX_{i}\wedge d\mu_{i}^{2}-
\nonumber\\
&-4\,L^{2}\:\sum_{i}X_{i}^{-2}\mu_{i}\,d\mu_{i}\wedge\left(d\varphi_{i}+\frac{1}{2\,L}\bar{A}_{i}\right)  \wedge\Hodge\bar{F}_{i}\:,
\end{align}
where $\Hodge$ is the Hodge dual with respect to the four-dimensional metric
$ds_{4}^{2}$, $\epsilon_{4}$ its volume form and $F$ is the four-form field
strength. The $\varphi_{i}$ are $2\pi$ periodic angular coordinates parametrizing
the four independent rotations on $S^{7}$. We will be interested in
considering the higher-dimensional interpretation of some of our solutions
using this uplift.\par
We shall work with a simplified version of this theory, where all dilatons take the same value, the so-called T${}^{3}$ model. In this case we set
\begin{equation}
\Phi_{a}=\sqrt{\frac{2}{3}}\,\phi\,,\qquad
\bar{F}_{1}=\sqrt{2}\,F^{1}\,,\qquad
\bar{F}_{2}=\bar{F}_{3}=\bar{F}
_{4}=\sqrt{\frac{2}{3}}\,F^{2}\,,\quad\label{eq:redefields}
\end{equation}
to obtain an action of the form:
\begin{equation}
\mathcal{S}=\frac{1}{\kappa}\int\! d^{4}x\:\,\sqrt{-g}\left(
    \frac{R}{2}-\frac{1}{2}\left(\partial\phi\right)^{2}
     +\frac{3}{L^{2}}\,\cosh\Big(\sqrt{\frac{2}{3}}\,\phi\Big)  -\frac{1}{4}\,e^{3\,\sqrt{\frac{2}{3}}\,\phi}\left(F^{1}\right)^{2}
    -\frac{1}{4}e^{-\sqrt{\frac{2}{3}}\,\phi}\left(F^{2}\right)^{2}\right),
\label{Lag}
\end{equation}
where $F_{\mu\nu}^{\Lambda}=\partial_{\mu}A_{\nu}^{\Lambda}-\partial_{\nu}A_{\mu
}^{\Lambda}$  \,$(\Lambda=1,2)$.\,
The field equations read
\begin{align}
&\partial_{\mu}\left(e^{3\sqrt{\frac{2}{3}}\,\phi}\,\sqrt{-g}\,F^{1\,\mu\nu}\right)=0\,,\qquad
\partial_{\mu}\left(e^{-\sqrt{\frac{2}{3}}\,\phi}\,\sqrt
{-g}\,F^{2\,\mu\nu}\right)=0\,,
\\[2ex]
&R_{\mu\nu}-\frac{1}{2}g_{\mu\nu}\,R  =e^{3\sqrt{\frac{2}{3}}\,\phi}\,
T_{\mu\nu}^{1}+e^{-\sqrt{\frac{2}{3}}\,\phi}\,T_{\mu\nu}^{2}+T_{\mu\nu}^{\phi}\,,
\\[2ex]
&\partial_{\mu}\left(\sqrt{-g}\,g^{\mu\nu}\,\partial_{\nu}\phi\right)+\frac{\sqrt{6}}{L^{2}}\,\sinh\Big(\sqrt{\frac{2}{3}}\,\phi\Big) =\frac{1}{2\sqrt{6}}\left( 3\,e^{3\sqrt{\frac{2}{3}}\,\phi
}\left(F^{1}\right)^{2}-e^{-\sqrt{\frac{2}{3}}\,\phi}\left(F^{2}\right)^{2}\right),
\end{align}
with
\begin{equation}
\begin{split}
T_{\mu\nu}^{\Lambda} &=F_{\mu\rho}^{\Lambda}\,F_{\nu}^{\Lambda}{}^{\rho}-\frac{1}{4}\,g_{\mu\nu}\,F_{\rho\sigma}^{\Lambda}F^{\Lambda\,\rho\sigma}\,,
\\[1.5ex]
T_{\mu\nu}^{\phi} &  =\partial_{\mu}\phi\,\partial_{\nu}\phi+g_{\mu\nu}\left(-\frac{1}{2}\left(\partial\phi\right)^{2}+\frac{3}{L^{2}}\,\cosh\Big(\sqrt{\frac{2}{3}}\,\phi\Big)\right)\:.
\end{split}
\end{equation}

\subsection{Supersymmetry}
The general formulae related to the supersymmetry transformations in  $\mathcal{N}=2$, $D=4$ supergravity with FI terms are given in Appendix \ref{app:susy}.
In what follows we shall restrict to the  T${}^3$ truncation of the STU model whose embedding in the ${\rm SO}(8)$-gauged maximal theory was outlined above.
This smaller T${}^3$ model can also be obtained from the more general class studied in \cite{Anabalon:2020pez} and labeled by a parameter $\nu$, by setting $\nu=-2$. This is also discussed in the aforementioned Appendix, where we also define our spinor conventions.
\paragraph{T{${}^3$} truncation.}
The fermionic variations \eqref{eq:susyvars}, once adapted to the T$^3$ model, read
\begin{align}
\delta\Psi^A_\mu\:=\;
    &\,\partial_\mu\epsilon^A+\frac{1}{4}\,{\omega_\mu}^{\!\!ab}\,\gamma_{ab}\,\epsilon^A- \frac{1}{2\,L}\left(\frac{1}{\sqrt{2}}\:A^1_\mu+\sqrt{\frac{3}{2}}\:A^2_\mu\right)i\left(\sigma^2\right)^A{\!\!}_B\;\epsilon^B+
\nonumber\\[-\jot]
    &+\frac{1}{8}\left(\frac{1}{\sqrt{2}}\:F^1_{\nu\rho}\,e^{\sqrt{\frac{3}{2}}\,\phi}
    +\sqrt{\frac{3}{2}}\:F^2_{\nu\rho}\,e^{-\frac{\phi}{\sqrt{6}}}\right)\,\gamma^{\nu\rho}\,\gamma_\mu\,\varepsilon^{AB}\:\epsilon_B+
\nonumber\\[\jot]
    &+\frac{1}{2}\;\mathcal{W}\;\gamma_\mu\,\delta^{AB}\,\epsilon_B\:,
\\[3ex]
\delta\lambda^{A}=
    &-\gamma^\mu\,\partial_\mu\phi\:\epsilon^A
    +\frac{1}{2\sqrt{2}}\left(-\sqrt{\frac{3}{2}}\:F^1_{\nu\rho}\,e^{\sqrt{\frac{3}{2}}\,\phi}
    +\frac{1}{\sqrt{2}}\:F^2_{\nu\rho}\,e^{-\frac{\phi}{\sqrt{6}}}\right)\,\gamma^{\nu\rho}\,\varepsilon^{AB}\,\epsilon_B-
    \nonumber\\
    &-\frac{1}{2\,L}\:\sqrt{\frac{3}{2}}\,\left(e^{-\sqrt{\frac{3}{2}}\,\phi}-e^{+\frac{\phi}{\sqrt{6}}}\right)\,\delta^{AB}\,\epsilon_B\;,
\end{align}
where the superpotential explicitly reads
\begin{equation}
\mathcal{W}=\frac{e^{-\sqrt{\frac{3}{2}}\,\phi}+3\,e^{\frac{\phi}{\sqrt{6}}}}{4\,L}\:.
\end{equation}
The above expressions coincide with those given in \cite{Duff:1999gh} once redefinitions \eqref{eq:redefields} are implemented in the latter.\par

With reference to the spinor conventions defined in Appendix \ref{app:susy}, we write the chiral spinors in terms of their real and imaginary parts:
\begin{equation}
\epsilon^{A}=\operatorname{Re}\epsilon^{A}+i\,\operatorname{Im}\epsilon^{A}\,,
\end{equation}
and  define the following complex spinors:
\begin{equation}
\chiR
=\operatorname{Re}\epsilon^{1}+i\,\operatorname{Re}\epsilon^{2}\,,
\qquad\quad
\chiI=\operatorname{Im}\epsilon^{1}+i\,\operatorname{Im}
\epsilon^{2}\,.
\end{equation}
As we shall see below, the Killing spinor equations can be written as first-order differential equations in each of them separately. The two, however, are not independent and we can solve the  Killing spinor equations in only one of them, for example $\chiR$.
Indeed, since in our spinor basis $\epsilon_A=\left(\epsilon^A\right)^*$, the Majorana spinors $\eM$ read:
\begin{equation}
\eM=\epsilon_A+\epsilon^A=2\,\operatorname{Re}\epsilon^A\:.
\end{equation}
The action of $\gamma^5=i\,\gamma^0\gamma^1\gamma^2\gamma^3$ on the above spinors gives
\begin{equation}
\gamma^5\,\eM=\epsilon_A-\epsilon^A=-2\,i\,\operatorname{Im}\epsilon^A\,,
\end{equation}
so that
\begin{equation}
\operatorname{Im}\epsilon^A=i\,\gamma^5\,\operatorname{Re}\epsilon^A\:,
\end{equation}
and therefore
\begin{equation}
\chiI\equiv \operatorname{Im}\epsilon^1+i\,\operatorname{Im}\epsilon^2
=i\,\gamma^5\left(\operatorname{Re}\epsilon^1+i\,\operatorname{Re}\epsilon^2\right)=i\,\gamma^5\,\chiR\:,
\label{XIXR}
\end{equation}
expressing the relation between $\chiR$ and $\chiI$.\par
Being $\epsilon^A$ and $\epsilon_A$ the chiral components of two Majorana spinors, the only freedom we have is to act on the solution by:
\begin{equation}
\chiR\rightarrow\,e^{i\Theta}\,\chiR
\quad\Rightarrow\quad
\chiI\rightarrow\,e^{i\Theta}\,\chiI\,,
\end{equation}
which is the ${\rm SO}(2)$ symmetry of the solution, this group being the one gauged in the $\mathcal{N}=2$ model. The corresponding transformation on the Weyl spinors $\epsilon^A$ is:
\begin{equation}
    \epsilon^A\,=S(\Theta)^A{}_B\;\mathring{\epsilon}^B\,,
\qquad\quad
    S(\Theta)^A{}_B=\begin{pmatrix*}[r]\cos(\Theta) & -\sin(\Theta)\cr \sin(\Theta) & \cos(\Theta)\end{pmatrix*}\,.
\label{eSe}
\end{equation}
From a solution $\chiR$ of the Killing spinor equations, one can extract the corresponding supersymmetry parameters as follows
\begin{equation}
\eM[1]=2\,\operatorname{Re}\chiR\,,
\qquad\quad
\eM[2]=2\,\operatorname{Im}\chiR\,,
\end{equation}
and
\begin{equation}
\begin{split}
\epsilon^1=\frac{\left(\mathds{1}-\gamma^5\right)}{2}\,\eM[1]=\left(\mathds{1}-\gamma^5\right)\,\operatorname{Re}\chiR\:,
\\[2.5ex]
\epsilon^2=\frac{\left(\mathds{1}-\gamma^5\right)}{2}\,\eM[2]=\left(\mathds{1}-\gamma^5\right)\,\operatorname{Im}\chiR\:.
\label{eq:epsfromchiR}
\end{split}
\end{equation}
The Killing spinor equations for $\chiR$ explicitly read
\begin{align}
0\:=\:
    &\:\partial_\mu\chiR+\frac{1}{4}\,{\omega_\mu}^{\!\!ab}\,\gamma_{ab}\,\chiR+ \frac{i}{2\,L}\,\left(\frac{1}{\sqrt{2}}\:A^1_\mu+\sqrt{\frac{3}{2}}\:A^2_\mu\right)\chiR-
\nonumber\\[-\jot]
    &-\frac{i}{8}\,\left(\frac{1}{\sqrt{2}}\:F^1_{\nu\rho}\,e^{\sqrt{\frac{3}{2}}\,\phi}
    +\sqrt{\frac{3}{2}}\:F^2_{\nu\rho}\,e^{-\frac{\phi}{\sqrt{6}}}\right)\gamma^{\nu\rho}\,\gamma_\mu\,\chiR+\frac{1}{2}\;\mathcal{W}\;\gamma_\mu\,\chiR\:,
\\[2.5ex]
0\:=\:
    &-\gamma^\mu\,\partial_\mu\phi\:\chiR
    -\frac{i}{2\sqrt{2}}\left(-\sqrt{\frac{3}{2}}\:F^1_{\nu\rho}\,e^{\sqrt{\frac{3}{2}}\,\phi}
    +\frac{1}{\sqrt{2}}\:F^2_{\nu\rho}\,e^{-\frac{\phi}{\sqrt{6}}}\right)\gamma^{\nu\rho}\,\chiR-
    \nonumber\\
    &-\frac{1}{2\,L}\:\sqrt{\frac{3}{2}}\,\left(e^{-\sqrt{\frac{3}{2}}\,\phi}-e^{+\frac{\phi}{\sqrt{6}}}\right)\chiR\;.
\end{align}
The corresponding conditions on  $\chiI$ are simply obtained by multiplying the above equations by \,$i\,\gamma^5$ from the left.

\bigskip

\section{Hairy soliton solutions}
\label{sol}
We can obtain soliton solutions of this theory, which generalize the soliton studied in \cite{Anabalon:2019tcy} by including a non-trivial scalar profile. These solutions are double analytic continuations of a particular case of the electrically charged black hole solutions that have been studied in \cite{Anabalon:2012ta, Anabalon:2017yhv, Anabalon:2020pez}. The charged planar black holes in this theory are a particular case of the charged STU model black holes \cite{Cvetic:1999xp}, which oxidize to spinning M2 branes.%
\footnote{%
%\sloppy
The STU model \cite{Duff:1995sm,Behrndt:1996hu,Behrndt:1997ny} is a $\mathcal{N}=2$ supergravity coupled to 3 vector multiplets and characterized, in~a suitable symplectic frame, by~the prepotential
\,${\mathcal{F}_\textsc{stu}(\mathcalboondox{X}^\Lambda)\:=-\frac{i}{4}\,\sqrt{\mathcalboondox{X}^0\,\mathcalboondox{X}^1\,\mathcalboondox{X}^2\,\mathcalboondox{X}^3}}$,\,
together with symmetric scalar manifold of the form \,${\mathscr{M}_\textsc{stu}=\left(\mathrm{SL}(2,\mathbb{R})/\mathrm{SO}(2)\right)^3}$\, spanned by the three complex scalars \,$z^i=\mathcalboondox{X}^i/\mathcalboondox{X}^0$\, ($i=1,2,3$); this model is in turn a consistent truncation of the maximal $\mathcal{N}=8$ theory in four dimensions with $\mathrm{SO}(8)$ gauge group \cite{Duff:1999gh,Andrianopoli:2013kya,Andrianopoli:2013jra,Andrianopoli:2013ksa}.
}\par
The vierbein and matter fields of the hairy soliton configuration can be obtained by means of a double Wick rotation
\begin{equation}
t\to i\,\varphi\,,
\qquad\quad
\varphi\to i\,t\,,
\qquad\quad
Q_\Lambda\to i\,Q_\Lambda\,,
\end{equation}
of the electrically charged planar black hole of \cite{Anabalon:2020pez} for $\nu=-2$ and read
\begin{align}
e^{0} & =\sqrt{\Upsilon(x)}\,dt,
\quad\,
e^{1}=\sqrt{\frac{\Upsilon(x)}{f(x)}}\;\eta\,dx,
\quad\,
e^{2}=\sqrt{\Upsilon(x)\,f(x)}\,d\varphi,
\quad\,
e^{3}=\sqrt{\Upsilon(x)}\,dz,
\\[1ex]
\phi&=\sqrt{\frac{3}{2}}\,\ln(x)\,,
\qquad
A^{1}=Q_{1}\left(  x^{-2}-x_{0}^{-2}\right)  \,d\varphi\,,
\qquad
A^{2}=Q_{2}\left(x^{2}-x_{0}^{2}\right)\,d\varphi\,,
\end{align}
with
\begin{align}\label{Fx}
\Upsilon(x)=\frac{4\,L^{2}\,x}{\left(x^{2}-1\right)^{2}\,\eta^{2}}\:,
\qquad\quad
f(x) =1+\frac{\eta^{2}\left(x^{2}-1\right)^{3}\left(3\,Q_{1}^{2}
-x^{2}\,Q_{2}^{2}\right)}{6\,L^{2}\,x^{2}}\;.\qquad
\end{align}
The conformal boundary of the metric is located at $x=1$, where the conformal factor of the metric has a pole of order 2. We remark that the boundary can be approached from the region with $x>1$ or from the region where $x<1$. Therefore, this form for the metric can represent two different spacetimes, one for $x$ in the range $x\in(0,1)$ and the other for $x$ in the range $x\in(1,\infty)$. The solutions with $x<1$ and $x>1$ are physically distinguished by the sign of the dilaton.\par
The canonical form of an asymptotically locally AdS$_{4}$ spacetime is achieved with the transformation
\begin{equation}
x=1\pm\left(  \frac{L^{2}}{\eta\,\rho}-\frac{L^{6}}{8\,\eta^{3}\rho^{3}}\right)
+\frac{L^{8}}{8\,\eta^{4}\rho^{4}}+O(\rho^{-5})\:,
\end{equation}
where the choice of sign depends on whether we are considering $x>1$ or $x<1$, which yields
\begin{align}
\Upsilon(x)  &  =\frac{\rho^{2}}{L^{2}}+O(\rho^{-2})\:,
\\[\jot]
g_{\varphi\varphi}  &  =\Upsilon(x)\,f(x)=\frac{\rho^{2}}{L^{2}}-\frac{\mu}{\rho
}+O(\rho^{-2})\:,
\\[\jot]
\mu & =\mp\frac{4\,L^{2}}{3\,\eta}\left(3\,Q_{1}^{2}-Q_{2}^{2}\right)\:.\label{mu}
\end{align}
As we shall see, $\mu$ is proportional to the energy of the configuration. The
expansion of the scalar field yields
\begin{align}
\phi=L^{2}\,\frac{\phi_{0}}{\rho}+L^{4}\,\frac{\phi_{1}}{\rho^{2}}+O(\rho^{-3})\:,
\end{align}
where
\begin{equation}
\phi_{0}   =\pm\frac{\sqrt{6}}{2\,\eta}\,,
\qquad
\phi_{1}=-\frac{\sqrt{6}}{4\,\eta^{2}}\,.
\end{equation}
We can regard these solitons as solutions with a boundary condition \,$\phi_{1}=-\frac{1}{\sqrt{6}}\,\phi_{0}^{2}$\, for the scalar field, which preserves conformal invariance. This boundary condition involves the cubic boundary conterterm corresponding to the triple-trace deformation, which is required, in the maximal theory, by supersymmetry \cite{Freedman:2016yue}.\par
We are interested in soliton solutions where the $\varphi$ circle contracts in the interior of the geometry, at some position $x_0$ such that
\begin{equation}
f(x_0)=0\:.
\end{equation}
If $x_0 <1$, we have a soliton with $x \in (x_0, 1)$, while if $x_0 >1$, we have a soliton with $x \in (1, x_0)$.\par
Regularity of the metric at $x=x_{0}$ requires $\varphi\in[0,\Delta]$ where
\begin{equation}
\Delta^{-1}
    =\left\vert \frac{1}{4\pi\,\eta}\:\frac{df}{dx}\right\vert_{{}_{x=x_{0}}}
    =\:\left\vert \frac{\eta\left(x_0^2-1\right)^{2}}{4\pi\,L^{2}\,x_0^3}\left(Q_1^2\left(1+2\,x_{0}^{2}\right) -Q_{2}^{2}\,x_{0}^{4}\right)\right\vert \:.
\end{equation}
We have normalized the gauge fields to vanish at $f(x_{0})=0$ to ensure their regularity.
Solutions with non-zero charges have net magnetic fluxes at infinity,
\begin{equation}
\begin{split}
\Phi_{\textsc{m}}^{1} & =\int F^{1}=\oint A^{1}=Q_{1}\,\Delta\left(1-x_{0}^{-2}\right)\equiv2\pi
L\,\psi_{1},
\\[\jot]
\Phi_{\textsc{m}}^{2}  &  =\int F^{2}=\oint A^{2}=Q_{2}\,\Delta\left(1-x_{0}^{2}\right)\equiv2\pi
L\,\psi_{2}\:.
\end{split}
\end{equation}
The scalar field induces a vev of an operator of conformal dimension one in the dual theory
\begin{equation}
\left\langle\mathcal{O}\right\rangle=\phi_{0}=
    \pm\frac{\sqrt{6}}{2}\:
    \frac{\pi\,x_{0}\left\vert\psi_{1}^{2}\left(1+2\,x_{0}^{2}\right)-\psi_{2}^{2}
    \right\vert}{\Delta}\:,
\end{equation}
and the source of the scalar vanishes on these solutions as it turns out to be proportional to ${\phi_{1}+\frac{1}{\sqrt{6}}\,\phi_{0}^{2}}$\,. The dual energy momentum tensor is given by \cite{Myers:1999psa,Anabalon:2015xvl,Astefanesei:2018vga,Astefanesei:2021ryn}
\begin{equation}
\left\langle T_{tt}\right\rangle =-\frac{\mu}{2\,\kappa\, L^{2}}\,,
\qquad
\left\langle T_{zz}\right\rangle =\frac{\mu}{2\,\kappa\,L^{2}}\,,
\qquad
\left\langle T_{\varphi\varphi}\right\rangle =-\frac{\mu}{\kappa\,L^{2}}\,.
\end{equation}
The gauge field gives a vev for the current in the boundary theory
\cite{Marolf:2006nd}
\begin{align}
\left\langle J^{\nu}_{1}\right\rangle  &  =\frac{\delta\mathcal{S}}{\delta A^{1}_{\nu}}=-\frac{1}{\kappa}\,N_{\mu}\,e^{3\sqrt{\frac{2}{3}}\,\phi}\,F^{1\,\mu\nu}\,\sqrt{\left\vert h\right\vert}=\frac{2\,Q_{1}}{\eta\,\kappa}\,\delta_{\varphi}^{\nu}\:,\label{GC1}
\\[\jot]
\left\langle J^{\nu}_{2}\right\rangle  &  =\frac{\delta\mathcal{S}}{\delta A^{2}_{\nu}}=-\frac
{1}{\kappa}\,N_{\mu}\,e^{-\sqrt{\frac{2}{3}}\,\phi}F^{2\,\mu\nu}\,\sqrt{\left\vert
h\right\vert}=-\frac{2\,Q_{2}}{\eta\,\kappa}\,\delta_{\varphi}^{\nu}\:, \label{GC2}
\end{align}
where $N_{\mu}$ is the outward pointing normal to the boundary metric
$h_{\mu\nu}=g_{\mu\nu}-N_{\mu}N_{\nu}$. The above calculation for the $J_\Lambda^{\nu}$ is valid for the
solution at $x<1$. Otherwise there is a flip of sign which can be traced back to the $N_{\mu}$. Note that the currents are proportional to $Q_\Lambda/\eta$.

\subsection{Existence of solitons}
From the bulk point of view, the solutions are parametrised by $Q_1$, $Q_2$ and $\eta$. It is simple to show that there are solitons for all non-zero values of the parameters. Indeed, non-zero parameters imply $f(x=1)=1$, $f(+\infty)<0$ and $f(0)<0$, so we must have $f(x_{0})=0$ for at least one value $x_0$ in $(0,1)$ and at least one value in $(1, \infty)$. Thus,  there are two soliton solutions, one for $x \in (x_- ,1)$, where $x_-$ is the largest root of $f$ for $x<1$, and one for $x \in (1, x_+)$, where $x_+$ is the smallest root of $f$ for $x>1$.\par
From the boundary point of view, however, it is more natural to parameterise solutions in terms of the boundary data we hold fixed: we can consider either fixed fluxes (Wilson loops), holding fixed $\psi_1$, $\psi_2$ and the period $\Delta$, or fixed charges, holding fixed $Q_1/\eta$, $Q_2/\eta$, and the period $\Delta$.%
\footnote{More precisely, this is a boundary condition of fixed boundary currents, which is the alternate boundary condition for the bulk gauge fields. We refer to this as fixed charges, although this language is more appropriate for the Euclidean continuation; the currents here are spacelike so they do not correspond to a physical charge density in the Lorentzian solution.} It turns out to be convenient to describe the fixed charge boundary conditions in terms of the rescaled parameters
\begin{equation}
q_{1}   \equiv\frac{\Delta^{2}}{4\pi^{2}L}\,\frac{Q_{1}}{\eta}\:, \qquad
q_{2}  \equiv\frac{\Delta^{2}}{4\pi^{2}L}\,\frac{Q_{2}}{\eta}\:.
\end{equation}
Solitons only exist for a range of values of $\psi_1, \psi_2$ or $q_1, q_2$.

\paragraph{Fixed fluxes.}
In order to study the existence of soliton solutions as functions of the boundary data $\psi_{1}, \psi_{2}$ and $\Delta$, we consider the bulk parameters as functions of the boundary data
\begin{equation} \label{Qfrompsi} Q_{1}=\frac{2\pi\,L\,\psi_{1}}{\Delta\,(1-x_{0}^{-2})}\,,
\qquad Q_{2}=\frac{2\pi\,L\,\psi_{2}}{\Delta\,(1-x_{0}^{2})}\,,
\qquad
\eta=\frac{x_{0}^{-1}\,\Delta}{\pi\left\vert \psi_{1}^{2}\left(1+2\,x_{0}^{2}\right)-\psi_{2}^{2}\right\vert }\,.
\end{equation}
Substituting these expressions into $f(x_0)$, we obtain
\begin{equation} \label{fxp}
    f(x_0)  = 1 + \frac{2}{3}\; \frac{3\left(1-x_0^2\right)^2 \psi_1^2 -\left(1-x_0^2\right)\left(3\,\psi_1^2 - \psi_2^2\right)}{x_0^2\,\big(2\left(1-x_0^2\right)\,\psi_1^2 -\left(3\,\psi_1^2 - \psi_2^2\right)\big)^2}\;.
\end{equation}
The solitons are thus parametrised by $x_0$ satisfying%
\footnote{To obtain this expression we multiplied out the denominator in the second term in $f$. This denominator is non-vanishing on the solution $x_0$ unless the numerator also vanishes, which only happens if $3\,\psi_1^2 - \psi_2^2 =0$; we will discuss this case in the next subsection.}
\begin{equation} \label{px}
P(x_{0})=4\,\psi_{1}^{4}\,x_{0}^{6}+2\,\psi_{1}^{2}\left(2\psi_{1}^{2}-2\psi_{2}^{2}+1\right)x_{0}^{4}+\left(\psi_{1}^{4}+\psi_{2}^{4}-2\psi_{1}^{2}\psi_{2}^{2}-2\psi_{1}^{2}-\frac{2}{3}\psi_{2}^{2}\right)x_{0}^{2}+\frac{2}{3}\psi_{2}^{2}=0.
\end{equation}
This is a cubic equation for $x_{0}^{2}$ and therefore there are at most three
different solutions. In fact, $P(x_0)$ is positive at large $x_0$,
$P(x_{0}=1)=(3\psi_{1}^{2}-\psi_{2}^{2})^{2}$ and $P(x_{0}=0)=\frac{2}{3}%
\psi_{2}^{2}$, so the cubic has either two or no roots for $x_0>0$. If it has two roots they are either both at $x>1$ or both at $x<1$.%
\footnote{This is consistent with the earlier analysis for fixed bulk parameters, since here a change the root also changes the values of the bulk parameters through \eqref{Qfrompsi}.}
The boundary between two roots and no roots occurs where $P(x_0)$ has a double root; solving $P(x_0)=0$ and $dP/dx_0 = 0$, we find that the boundary is given by
\begin{equation}
36\,\psi_1^6 - 36\,\psi_2^6 +108\, \psi_1^4 \psi_2^2 - 108\, \psi_1^2 \psi_2^4 - 141\, \psi_1^4 -258\, \psi_1^2 \psi_2^2 + 75\, \psi_2^4 +132\, \psi_1^2 - 52\, \psi_2^2 +12 =0.
\end{equation}
As \eqref{px} is a simple cubic equation for $x_0^2$, we have been able to solve it analytically and find that indeed there are two real roots characterizing two solutions, this can be found in the Appendix \ref{app:roots}. Hence we see that, as happens in the Einstein-Maxwell system \cite{Kastor:2020wsm, Anabalon:2021tua}, there is a range of parameters where there are two different soliton solutions, which coalesce at the boundary of the parameter range.

\paragraph{Fixed charges.}
For fixed charges the situation is more intricate. We determine $\eta$ by
\begin{equation}
\eta =\frac{3}{2\pi}\:\Delta\:\frac{\left\vert x_{0}^{4}\,q_{2}^{2}-2\,x_{0}^{2}\,q_{1}^{2}-q_{1}^{2}\right\vert }{x_{0}\left(  x_{0}^{2}-1\right) \left(x_{0}^{2}\,q_{2}^{2}-3\,q_{1}^{2}\right)}\;,
\end{equation}
and the polynomial to be solved is
\begin{equation}
f(x_{0})=1+\frac{\left(  x_{0}^{4}\,q_{2}^{2}-2\,x_{0}^{2}\,q_{1}^{2}-q_{1}^{2}\right)^{4}}{x_{0}^{6}\left(x_{0}^{2}-1\right)\left(3\,q_{1}^{2}-q_{2}^{2}\,x_{0}^{2}\right)^{3}}\;\frac{3^{3}}{2}=0\;.
\label{Pol}
\end{equation}
This polynomial is of order $8$ in $x_{0}^{2}$\,. We find that by setting
\begin{equation}
q_{2}^{2}=\frac{2}{27}\frac{\left(x_{0}^{2}-1\right)  }{x_{0}^{4}}\,\lambda^{3}+q_{1}^{2}\,\frac{\left(  1+2\,x_{0}^{2}\right)}{x_{0}^{4}}\;,
\label{pol2}
\end{equation}
all the dependence on $x_{0}^{2}$ drops out from (\ref{Pol}). One is thus left
with four quadratic equations for $x_{0}^{2}$ of the form (\ref{pol2}), one for
each value of the parameter $\lambda^3$ given by
\begin{equation}
8\,\lambda^{9}(\lambda^{2}-1)(\lambda^{2}+\lambda+1)+4\times3^{4}\,q_{1}^{2}\,\lambda^{6}-2\times3^{7}\,q_{1}^{4}\,\lambda^{3}+3^{9}\,q_{1}^{6}=0\:.
\label{pol3}
\end{equation}
We can also view this equation as a cubic for $q_{1}^{2}$, with three solutions
\begin{equation}
q_{1}^{2}=\frac{2}{3^3}\left(\lambda^{3}-\lambda^{4}\,e^{\frac{2\pi i}{3}k}\right)\:,
\end{equation}
with $k=0,1,2$. Real solutions exists only for $k=0$. Thus, for given $q_1, q_2$, we can determine $\lambda$ from $q_{1}^{2}=\frac{2}{3^3}\,\left(\lambda^{3}-\lambda^{4}\right)$ and then determine $x_0$ from \eqref{pol2}.\par
\begin{figure}[H]
\centering
\includegraphics[scale=0.4]{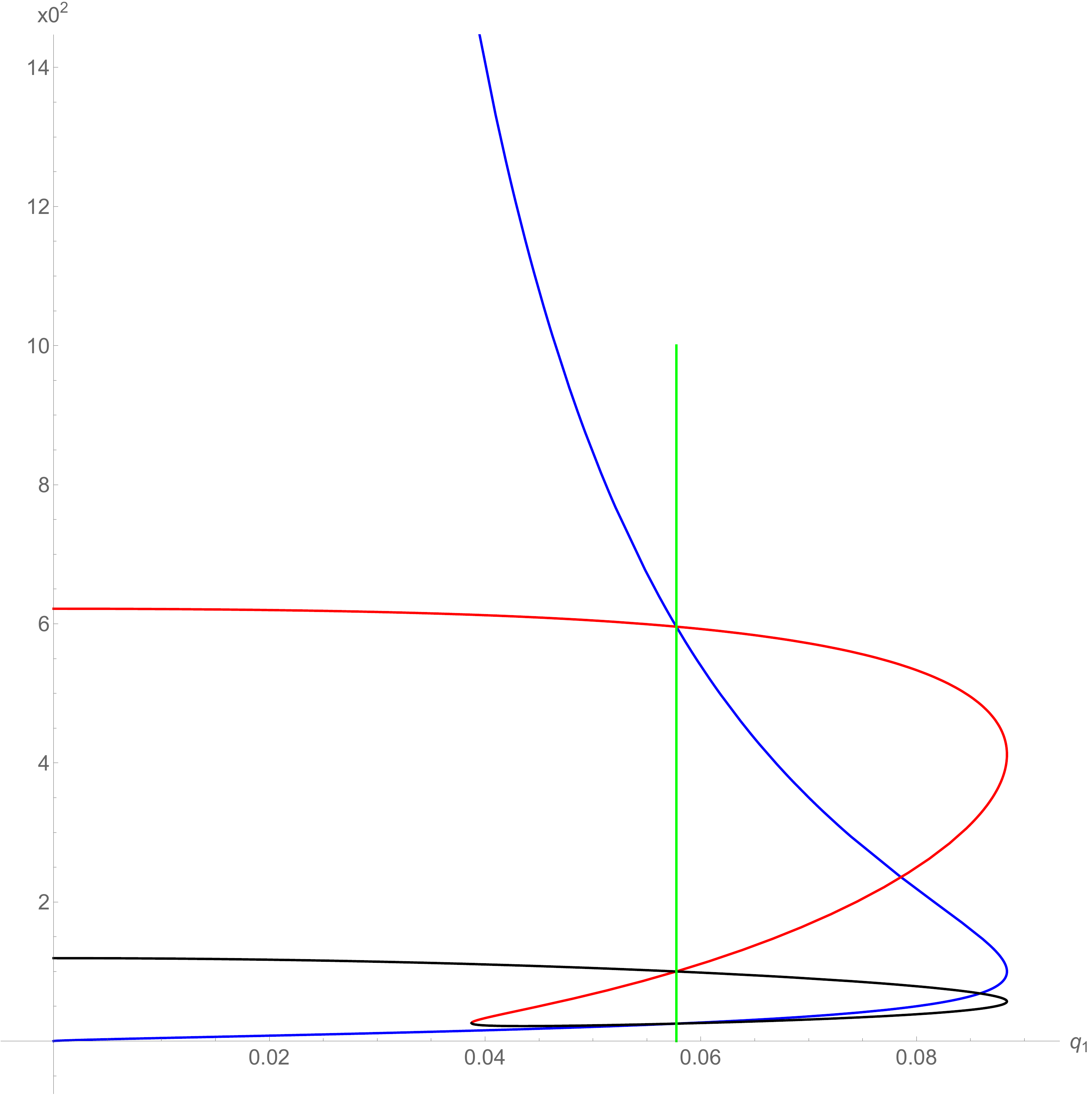}
\caption{The square $x_0^2$ of the roots of (\ref{Pol}) in the $y$-axis vs.\ the ``rescaled charge'' $q_1$ in the $x$-axis. The blue line shows the location of the supersymmetric solitons, where $q_2$ is determined as a function of $q_1$ by ${q_2=-\sqrt{3}\,q_1}$ (see Section \ref{susy}). The red and black lines are the roots of (\ref{pol2}) plotted for fixed ${q_2 = -0.1}$. The green line indicates the value of $q_1$ that that satisfies the susy condition $q_1=-\frac{1}{\sqrt{3}}\,q_2$ at $q_2=-0.1$. As expected, this intersects the red and black lines where they intersect the blue line: these are the supersymmetric solitons for fixed $q_2=-0.1$. There is also an intersection at $x_0=1$, where also the black and red roots intersect; we will see below that this corresponds to non-supersymmetric solutions with zero scalar.}%
\label{fig:fig1}
\end{figure}
\begin{figure}[H]
\centering
\includegraphics[scale=0.4]{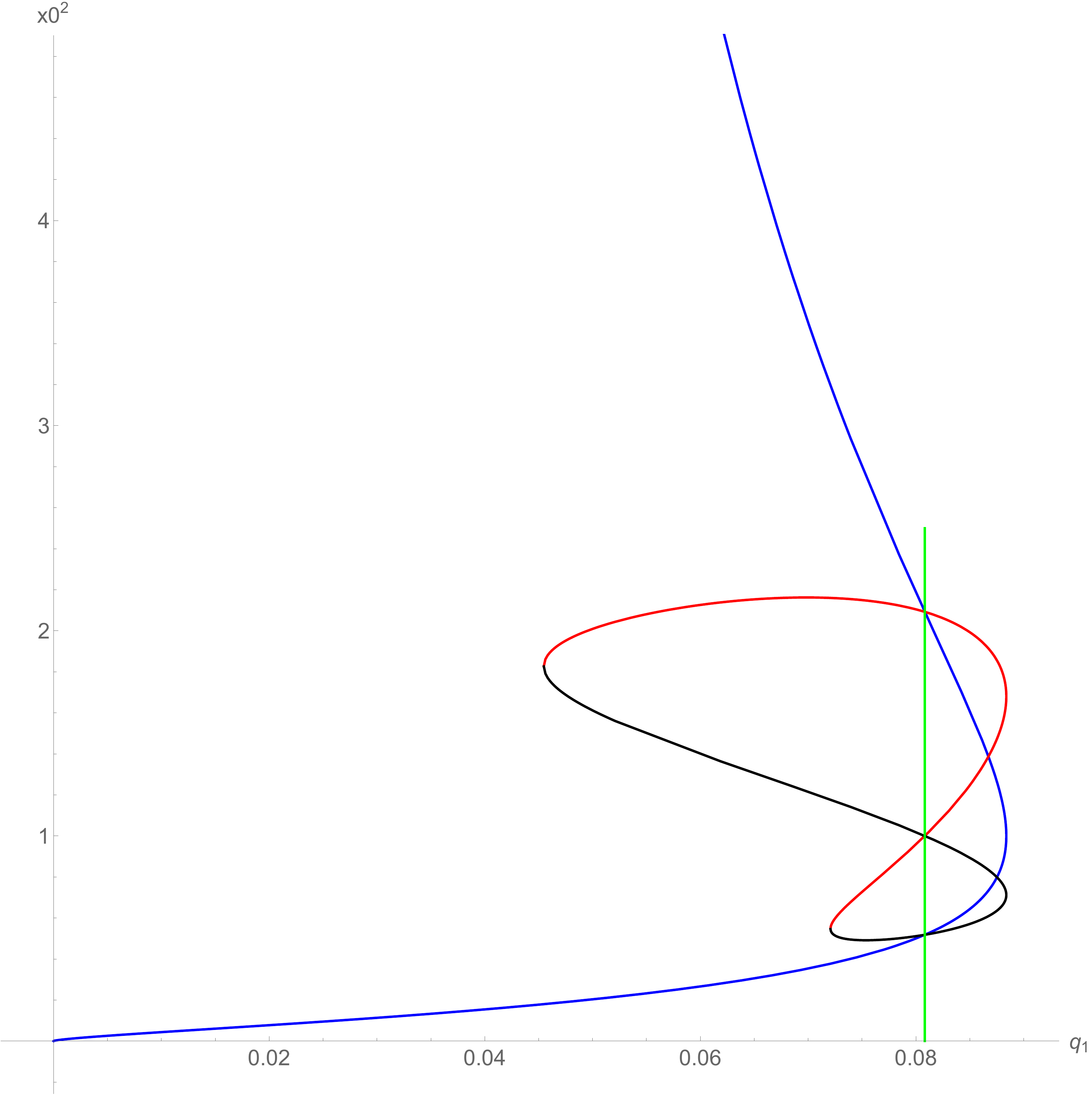}
\caption{\sloppy%
The square $x_0^2$ of the roots of (\ref{Pol}) in the $y$-axis vs.\ the ``rescaled charge'' $q_1$ in the $x$-axis. The blue line is the same as in the previous plot - it shows the location of the supersymmetric solitons, where $q_2$ is determined as a function of $q_1$ by ${q_2= - \sqrt{3}\,q_1}$ (see Section \ref{susy}). The red and black lines are the roots of (\ref{pol2}) plotted for fixed ${q_2 = -0.14}$. We see that for this value of $q_2$, there is both a lower and an upper bound on $q_1$ for the existence of solitons. The green line  indicates the value of $q_1$ that  satisfies the susy condition $q_1= - \frac{1}{\sqrt{3}}\, q_2$ at $q_2=-0.14$. As expected, this intersects the red and black lines where they intersect the blue line: these are the supersymmetric solitons for fixed $q_2=-0.14$. There is also an intersection at $x_0=1$, where also the black and red roots intersect; we will see below that this corresponds to non-supersymmetric solutions with zero scalar.}%
\label{fig:fig2}
\end{figure}

The
number of solitons that might exists for every value of the parameters
$\left( q_{1}^{2},\,q_{2}^{2}\right)  $ is not obvious. We find that there could be between zero and four solutions, as can be seen from the plots of the roots in Figures \ref{fig:fig1} and \ref{fig:fig2}. The analysis of the supersymmetric cases is much more simple and we found that there are $2$ superymmetric solitons for the same boundary conditions, both solutions featuring the same energy and free energy.

In Figure \ref{fig:fig1} it is possible to see that, for this region of the phase space, there are two solitons at low $q_1$ and there could be four solitons for larger values of $q_1$. The supersymmetric configurations are embedded in the region of the phase space where there are four solitons. In Figure \ref{fig:fig2} we show a region of the phase space where the structure of the roots seems to be different but the number of solutions is the same, and again there are two supersymmetric configurations. Let us remark that the structure of the solution space is actually very different than for pure Einstein-Maxwell-AdS system, where there are only two solitons at each value of the charge \cite{Anabalon:2021tua}.

\subsection{Relation to earlier solutions}
In \cite{Anabalon:2021tua}, solutions were found in a theory with a single gauge field, obtained by compactification from $D=11$ supergravity setting \,$\bar A_i = \frac{1}{2}\,A$  \,($i = 1, \ldots 4$)\, and \,$\Phi_a =0$ \,($a=1,2,3$). The theory we have considered here thus reduces to the theory studied there if we set \,$A^1 = \frac{1}{\sqrt{3}} A^2 = \frac{1}{2 \sqrt{2}} A$\, and \,$\phi =0$.\par
We would like to understand the relation between the solutions obtained here and the solutions in \cite{Anabalon:2021tua}. This is complicated by the fact that the coordinate system adopted here is adapted to the scalar field, and that the functional form of $A^1$ and $A^2$ is different. Suppose we make the change of coordinates
\begin{equation}
x^2 = 1 - \frac{\alpha}{r}\:,
\end{equation}
so that the boundary lies at $r \to \infty$, and $\alpha$ is a parameter to be determined. We also define $r_0$ by
\begin{equation}
x_0^2 = 1 - \frac{\alpha}{r_0}\:.
\end{equation}
Then, the vierbein and matter fields are \begin{equation}
\begin{split}
&e^{0}=\sqrt{\Upsilon(r)}\,dt,\quad
e^{1}=\sqrt{\frac{\Upsilon(r)}{f(r)}}\,\frac{\alpha\,\eta}{2\,r^{3/2}\,(r - \alpha)^{1/2} }\,dr,\quad
e^{2}=\sqrt{\Upsilon(r)\,f(r)}\,d\varphi,\quad
e^{3} =\sqrt{\Upsilon(r)}\,dz,
\\[1ex]
&\phi=\sqrt{\frac{3}{8}}\,\ln\left(1-\frac{\alpha}{r}\right),
\quad\;
A^{1}=\frac{Q_1\,\alpha\left( \frac{1}{r} - \frac{1}{r_0} \right)}{\left(1- \frac{\alpha}{r}\right)\left(1 - \frac{\alpha}{r_0}\right)}\,d\varphi,
\quad\;
A^{2}=-Q_2\,\alpha \left( \frac{1}{r} - \frac{1}{r_0} \right) d \varphi,
\end{split}
\end{equation}
with
\begin{equation}
\Upsilon(r)  =\frac{4\,L^{2}\sqrt{1-\frac{\alpha}{r}} \:r^2}{\alpha^2\,\eta^{2}}\,,
\qquad\quad
f(r) =1-\frac{\eta^{2}\,\alpha^3 \left( 3\,Q_{1}^{2}%
-Q_{2}^{2} + Q_2^2\,\frac{\alpha}{r} \right)}{6\,L^{2}\,r^3\left(1 - \frac{\alpha}{r}\right)}\,.\qquad
\end{equation}
We see that we would obtain a vanishing scalar and the same functional form for $A^1$ and $A^2$ by taking $\alpha \to 0$, but this seems like a singular limit. However, we can obtain a well-behaved solution in this limit by scaling the parameters of the solution appropriately. If we take $\eta \to \infty$ with $\eta\,\alpha$ fixed, the vierbein has a finite limit and $\Upsilon \to \frac{4\,L^2 r^2}{\alpha^2 \eta^2}$. Let's choose coordinates so that $\eta\,\alpha \to 2L^2$ in the limit, so that $\Upsilon \to \frac{r^2}{L^2}$, reproducing the usual form. Now recall that the boundary  currents are proportional to $\tilde Q_1 = Q_1/ \eta$,\, $\tilde Q_2 = -Q_2/\eta$, so the physical limit with fixed vevs is $\eta \to \infty$, holding $\tilde Q_1$, $\tilde Q_2$ fixed. There is a final problem in $f(r)$, where, in order to make the second term finite in the limit, we must take \,$3\,\tilde Q_1^2 - \tilde Q_2^2 \to 0$, with $\mu = \frac{4\, \eta\,L^2}{3} (3\,\tilde Q_1^2 - \tilde Q_2^2)$ fixed. This is the expected limit where we set the two gauge fields equal. Indeed, this limit gives
\begin{equation}
\phi=0\,,
\qquad\quad
A^1 = \frac{1}{\sqrt{3}}\,A^2
= 2\,L^2 \tilde Q_1 \left( \frac{1}{r} - \frac{1}{r_0} \right) d \varphi,
\label{eq:limitold}
\end{equation}
and
\begin{align}
e^{0} =\frac{r}{L}\,dt\,,\qquad
e^{1}  = \frac{dr}{\sqrt{f_0(r)}}\,,\qquad
e^{2} =\sqrt{f_0(r)}\,d\varphi\,,\qquad
e^{3} =\frac{r}{L}\,dz\,,\qquad
\end{align}
with
\begin{equation}
f_0(r)= \frac{r^2}{L^2}\,f(r)= \frac{r^2}{L^2} - \frac{\mu}{r} - \frac{8\,L^4\,\tilde{Q}_1^2}{r^2}\:,
\end{equation}
which matches the solution in \cite{Anabalon:2021tua} with $Q = 2 \sqrt{2}\,L^2\,\tilde{Q}_1$. Thus, as expected, the solutions in \cite{Anabalon:2021tua} arise as limit of the solutions considered here, when we choose parameters such that the scalar field vanishes and the gauge fields are equal. \par
We are interested in studying the set of solitons obtained as a function of the boundary conditions. In particular, one might wonder whether it is possible to find new hairy configurations, from our more general model, satisfying the same boundary conditions as the previous solutions.
% If we choose boundary conditions where the earlier solutions satisfy the boundary conditions, are they then the only solutions or do we get further hairy solutions satisfying the same boundary conditions from our more general family?
\par
For fixed flux boundary conditions, the previous solutions satisfy boundary conditions with \,$3\,\psi_1^2 = \psi_2^2$, see \eqref{eq:limitold}. With the latter condition, the polynomial \eqref{px} becomes
\begin{equation}
    P(x_0) = 2\,\psi_1^2 \left(1-x_0^2\right)^2\left(1+2\,\psi_1^2\,x_0^2\right)\,,
\end{equation}
so it seems that the only real solution is $x_0=1$. Thus, as $3\,\psi_1^2 - \psi_2^2 \to 0$, we need $x_0 \to 1$, which is achieved by taking the limit $\eta \to \infty$ in which these solutions reduce to the previous solutions. That is, the only solutions with these boundary conditions are the ones with $\phi=0$ obtained previously.\par
The above conclusion is correct, but is worth to analyse this limit slightly more carefully, as there is a subtlety in the relation between $f(x_0)$ and $P(x_0)$. As previously noted, $P$ is obtained from $f$ by multiplying out the denominator in the second term in \eqref{fxp}, but when $3\,\psi_1^2 - \psi_2^2 =0$ and $x_0=1$ this denominator vanishes, so we should be more careful about this operation. First, we set \,$3\,\psi_1^2 - \psi_2^2 = \epsilon\,\psi_1^2$\, and \,$x_0^2 = 1-\epsilon\,y_0$. Then, as $\epsilon \to 0$\,,
\begin{equation}
  f(y_0)  = 1 + \frac{2}{3\,\psi_1^2} \:\frac{3\,y_0^2 - y_0}{(2\,y_0 - 1)^2}\:.
\end{equation}
This indeed has two solutions for $y_0$, when $\psi_1 < \frac{1}{\sqrt{6}}$. We also see that, as $\epsilon \to 0$ for fixed $\Delta$, \,$\eta \to \infty$ with fixed $\tilde Q_1$, $\tilde Q_2$, so we are in the limit where the solutions here reduce to the solutions of \cite{Anabalon:2021tua}. Thus, for the boundary conditions with fixed flux and $3\,\psi_1^2 = \psi_2^2$, the only soliton solutions are the ones found in \cite{Anabalon:2021tua}. \par\smallskip
For fixed charge, the situation is more interesting. Expressing $f$ in terms of the charge parameters $q_1$, $q_2$, we have
\begin{equation}
    f(x_0)  = 1 - \frac{3^3}{2}\:\frac{ \left(q_2^2 - 3\,q_1^2 - 2\left(1-x_0^2\right)\left(q_2^2 - q_1^2\right) + \left(1-x_0^2\right)^2\,q_2^2 \right)^4}{x_0^6\left(1-x_0^2\right) \big(3\,q_1^2 - q_2^2 + q_2^2 \left(1-x_0^2\right)\big)^3}\;,
\end{equation}
and
\begin{equation}
\eta =  \frac{3\,\Delta}{2 \pi}\:\frac{ \left|\left(q_2^2-3\,q_1^2\right) - 2 \left(1-x_0^2\right) \left(q_2^2 - q_1^2\right) + \left(1-x_0^2\right)^2 q_2^2 \right|}{x_0 \left(1-x_0^2\right) \left(3\,q_1^2 - q_2^2 + q_2^2 \left(1-x_0^2\right)\right) }\:.
\end{equation}
The solutions of \cite{Anabalon:2021tua} satisfy the boundary conditions for $q_2^2=3\,q_1^2$, and can be obtained from these solutions again by scaling $x_0^2 \to 1$\, as \,$q_2^2 - 3\,q_1^2 \to 0$. But in this case there are also solutions with \,$q_2^2 - 3\,q_1^2 = 0$\, for \,$x_0 \neq 1$, setting
\begin{equation}
    f(x_0)  = 1 - \frac{q_1^2}{2\,x_0^6} \left(1+3\,x_0^2\right)^4 = 0\:,
\end{equation}
and
\begin{equation}
\eta =  \frac{\Delta}{2 \pi}\:\frac{ 1+3\,x_0^2}{x_0\left(1-x_0^2\right)}\:.
\end{equation}
Indeed, for $q_1 = - \frac{1}{\sqrt{3}}\, q_2$, these are the supersymmetric solutions which are the main focus of our subsequent interests.

Thus, for the boundary conditions in the fixed charge case, we have both the  supersymmetric hairy solutions and the solutions with $\phi=0$ obtained in \cite{Anabalon:2021tua}. The latter are not supersymmetric, except at the maximum value of the period $\Delta$. These non-supersymmetric solutions correspond to the additional intersection of the green line with the red and black lines at $x_0=1$ in the plots in Figures \ref{fig:fig1} and \ref{fig:fig2}.

\sloppy
At the maximum value of $\Delta$, where the previous solutions become supersymmetric, they coincide with the supersymmetric hairy solitons found here. In \cite{Anabalon:2021tua}, it was found that fixed charge boundary conditions gave the supersymmetric solution for ${\Delta \varphi = \pi \sqrt{\frac{L^3}{Q}}}$.\, As \,$Q = 2 \sqrt{2}\,L^2\,\tilde Q_1$, this corresponds to $\Delta = \pi \sqrt{\frac{\eta\,L}{2 \sqrt{2}\,Q_1}}$, which gives $q_1= \frac{1}{8 \sqrt{2}}$, where the only real solution of \,$f(x_0)=0$\, is \,$x_0^2=1$, implying $\eta \to \infty$, so that we are again in the limit where these solutions reduce to the previous ones. This is illustrated in Figure \ref{fig:fig6}.

\section{Constant flux solutions}
\label{domw}
In addition to these soliton configurations, for the fixed flux boundary conditions, we can obtain a solution satisfying the latter by adding constant Wilson lines to a simple background. As in \cite{Anabalon:2021tua}, we can obtain such solutions by adding the Wilson lines to Poincar\'{e}-AdS.

There is now an additional possibility: we can take domain wall solutions with a non-trivial scalar profile and add Wilson lines. The relevant planar domain walls can be obtained from our general solution by setting $Q_1=Q_2=0$. We then have $f(x)=1$, and the $\varphi$ circle is non-contractible in the interior of the spacetime, but the solution still has a non-trivial scalar profile. The domain wall can be obtained as a limit of the solutions originally obtained in \cite{Cvetic:1999xp}. The domain wall configurations have a curvature singularity at $x=0$, where the dilaton blows up, and the classical gravity description will break down. Both Poincar\'e-AdS and the domain wall would be supersymmetric solutions if we had periodic boundary conditions on the $\varphi$ circle and no flux.\par
Both Poincar\'{e}-AdS and the domain wall have vanishing gauge field strengths, but they can be made to satisfy the fixed flux boundary conditions by adding constant Wilson lines around the $\varphi$ circle, %
\begin{equation}
    A^1 = \frac{2\pi\,L\,\psi_1}{\Delta}\:, \qquad\quad A^2 = \frac{2 \pi\,L\, \psi_2}{\Delta}\:.
\end{equation}
These imply that we have a shift on the $S^7$ as we go around the $\varphi$ circle.

\subsection{Quantization of fluxes}
A point that was missed in \cite{Anabalon:2021tua} is that these fluxes must be quantized to make the Poincar\'e-AdS solution regular everywhere. Poincar\'e-AdS has a coordinate singularity on the horizon at $z = \infty$ in the standard coordinates, where the metric is%
\footnote{Note that the $z$ coordinate here is unrelated to the one appearing in our soliton solutions earlier; $z$ here is a radial coordinate analogous to $x$ earlier, but with the boundary at $z=0$.}
\begin{equation}
    ds^2 = \frac{L^2}{z^2} \left(dz^2 + dy^a dy_a\right)\,.
\end{equation}
To satisfy our boundary conditions, we must periodically identify one of the boundary coordinates. This quotient of AdS has fixed points, which lie in the horizon at $z = \infty$.

To identify the fixed points, it is convenient to embed AdS in a higher
dimensional spacetime with signature $\left(  -,-,+,+,+\right)$, with the constraint
\begin{equation}
-\left(  X^{-1}\right)  ^{2}+X^{a}X_{a}+\left(  X^{3}\right)  ^{2}
=-L^{2}\,.
\end{equation}
The higher dimensional embedding coordinates are related to the Poincar\'e coordinates by
\begin{equation}
X^{-1} =\frac{1}{2\,z}\left(  z^{2}+L^{2}+y^{a}y_{a}\right), \quad\;\;
X^{a} =\frac{L\,y^{a}}{z}\,,\quad\;\;
X^{3} =\frac{1}{2\,z}\left(  z^{2}-L^{2}+y^{a}y_{a}\right),
\end{equation}
with $a=0,1,2$\, and \,$\eta_{ab}=\left(  -,+,+\right)$.

Suppose we pick $y^{1}$ to be periodic. In terms of the embedding coordinates we have
\begin{equation}
\frac{\partial}{\partial y^{1}}   =L^{-1}\left( X^{-1}-X^{3}\right)  \frac{\partial}{\partial X^{1}}
+L^{-1}X^{1}\left(  \frac{\partial}{\partial X^{-1}}+\frac{\partial}{\partial
X^{3}}\right),
\end{equation}
so the Killing vector has fixed points when
\,$X^{-1}-X^{3}=0$\,  and \,$X^{1}=0$\, (the former corresponds to $z = \infty$ in the Poincar\'e coordinates).  If we simply made a periodic identification in $y^1$, we would have singularities at these fixed points. Since we have added Wilson lines, the identification in $y^1$ is made with a shift on the $S^7$. The identification is then freely acting on AdS${}_4 \times S^7$. At the fixed points of the action on AdS${}_4$, we have a quotient of the sphere generated by
\begin{equation}
\bar \varphi_1 \sim \bar \varphi_1 + 2\pi\,\frac{\psi_1}{\sqrt{2}}\,, \qquad\quad
\bar \varphi_{i \neq 1} \sim \bar \varphi_{i \neq 1} + 2\pi\, \frac{\psi_2}{\sqrt{6}}\,.
\end{equation}
This seems to lead to a completely regular solution. A formal discussion of this identification is given in Appendix \ref{S7id}.

However, the point which was missed in \cite{Anabalon:2021tua} is that the coordinates $\varphi_i$ are $2\pi$~periodic on the sphere, so for the quotient to be well-behaved we need this identification to involve shifts in $\varphi_i$ which are rational multiples of $2\pi$. That is, we have Poincar\'{e}-AdS solutions for the fixed flux boundary conditions when $\psi_1/\sqrt{2}$, $\psi_2/\sqrt{6}$ are rational numbers. The somewhat surprising consequence is that these solutions exist only for a dense subset of the possible fixed flux boundary conditions.

Formally, we can apply a similar analysis to the domain wall solution. The periodicity in the $\varphi$ direction is imposed by an appropriate identification on the domain wall, and the presence of the Wilson lines implies that this identification involves a shift on the sphere. The identification on the domain wall has fixed points at $x=0$, where we have a quotient on the sphere, and imposing regularity of this quotient would restrict us to rational fluxes. However, in the case of the domain wall, the solution is already singular at $x=0$ because of a divergence in the dilaton, so it is not clear whether it is necessary to impose regularity of the quotient of the sphere for this case.

\section{Supersymmetric solutions}
\label{susy}
The solutions will preserve some supersymmetry when
\begin{equation}
Q_{1}=-\frac{1}{\sqrt{3}}\,Q_{2}
\quad\Longrightarrow\quad
\mu=0\;,
\end{equation}
as can be obtained by explicitly constructing the Killing spinors.%
\footnote{%
Note that supersymmetry implies the vanishing of the energy parameter $\mu$ (see \eqref{mu}) as expected, but the converse is not true; in fact, the solution is not supersymmetric for $Q_1 = \frac{1}{\sqrt{3}}\,Q_{2}$, even though this also gives $\mu=0$.}
In this case the metric function $f(x)$ drastically simplifies, giving
\begin{equation}
f(x)=1-\frac{\left(x^{2}-1\right)^{4}}{x^{2}}\,\frac{\eta^{2}\,Q_{2}^{2}}{6\,L^{2}}\:,
\end{equation}
and we find that the Killing spinors are
\begin{alignat}{2}
\chi_{{}_{\text{R}_{(1)}}} &  =e^{i\,\omega\,\varphi}\left(
\begin{array}{@{}c@{}}
\alpha_{-}(x) \\
0\\
0\\
-i\,\alpha_{+}(x)
\end{array}
\right),
\qquad
\chi_{{}_{\text{R}_{(2)}}}  &=e^{i\,\omega\,\varphi}
\left(
\begin{array}{@{}c@{}}
0\\
\alpha_{+}(x) \\
-i\,\alpha_{-}(x) \\
0
\end{array}
\right),
\\[2ex]
\chi_{{}_{\text{I}_{(1)}}} &  =e^{i\,\omega\,\varphi}
\left(
\begin{array}{@{}c@{}}
0\\
\alpha_{-}(x) \\
-i\,\alpha_{+}(x) \\
0
\end{array}
\right),
\qquad
\chi_{{}_{\text{I}_{(2)}}} &=e^{i\,\omega\,\varphi}\left(
\begin{array}{c}
-\alpha_{+}(x) \\
0\\
0\\
i\,\alpha_{-}(x)
\end{array}
\right),
\end{alignat}
being \,$\chi_{{}_{\text{I}_{(k)}}}\!=i\,\gamma^5\chi_{{}_{\text{R}_{(k)}}}$\, (see \eqref{XIXR})\,  and
\begingroup
\setlength{\abovedisplayskip}{4pt plus 2pt minus 4pt}
\begin{equation}
\alpha_{\pm}(x) =\frac{x^{1/4}}{\left( x^{2}-1\right)^{1/2}}\,\sqrt{1\pm f(x)^{1/2}}\,,
\qquad\quad
\omega=-\frac{\pi}{\Delta}\,.\qquad
\end{equation}
\endgroup
Hence, the spinors are antiperiodic. The chiral spinors can be reconstructed
as
\begin{equation}
\begin{split}
{\epsilon}_{(k)}^{1}=\operatorname{Re}\chi_{{}_{\text{R}_{(k)}}}+i\,\operatorname{Re}\chi_{{}_{\text{I}_{(k)}}}
=\left(\mathds{1}-\gamma^5\right)\,\operatorname{Re}\chi_{{}_{\text{R}_{(k)}}}\,,
\\[1.5ex]
{\epsilon}_{(k)}^{2}=\operatorname{Im}\chi_{{}_{\text{R}_{(k)}}}+i\,\operatorname{Im}\chi_{{}_{\text{I}_{(k)}}}
=\left(\mathds{1}-\gamma^5\right)\,\operatorname{Im}\chi_{{}_{\text{R}_{(k)}}}\,.
\end{split}
\end{equation}
Explicitly, they read
\begin{align}
{\epsilon}_{(1)}^{1}  &  =
\left(
\begin{array}[c]{r@{}l}
%[c]{c}%
&\cos(\omega\varphi)\,\alpha_{-}(x) \\
i\,&\cos(\omega\varphi)\,\alpha_{-}(x)\\
i\,&\sin(\omega\varphi)\,\alpha_{+}(x)\\
&\sin(\omega\varphi)\,\alpha_{+}(x)
\end{array}
\right),
\qquad
{\epsilon}_{(1)}^{2}=
\left(
\begin{array}[c]{@{}r@{}l}
&\sin(\omega\varphi)\,\alpha_{-}(x)\\
i\,&\sin(\omega\varphi)\,\alpha_{-}(x)\\
-i\,&\cos(\omega\varphi)\,\alpha_{+}(x)\\
-&\cos(\omega\varphi)\,\alpha_{+}(x)
\end{array}
\right),\label{KS1}
\\[2.75ex]
{\epsilon}_{(2)}^{1}  &  =
\left(
\begin{array}{@{}r@{}l}
-i\,&\cos(\omega\varphi)\,\alpha_{+}(x)\\
&\cos(\omega\varphi)\,\alpha_{+}(x)\\
&\sin(\omega\varphi)\,\alpha_{-}(x)\\
-i\,&\sin(\omega\varphi)\,\alpha_{-}(x)
\end{array}
\right),
\qquad
{\epsilon}_{(2)}^{2}=
\left(
\begin{array}{@{}r@{}l}
-i\,&\sin(\omega\varphi)\,\alpha_{+}(x)\\
&\sin(\omega\varphi)\,\alpha_{+}(x)\\
-&\cos(\omega\varphi)\,\alpha_{-}(x)\\
i\,&\cos(\omega\varphi)\,\alpha_{-}(x)
\end{array}
\right),
\label{KS2}
\end{align}
and indeed satisfy
\begin{equation}
\gamma^{5}\,{\epsilon}_{(k)}^{A}=-{\epsilon}_{(k)}^{A}\:.
\end{equation}
A more general basis can be obtained by acting on $\epsilon^A$ by the transformation \eqref{eSe}. Using this redefinition, in the special case \,$\alpha_-=0=\omega$\, (corresponding to the domain wall solutions we discuss below) although two of the spinors in \eqref{KS2} vanish, we still have two independent solutions $ \epsilon^A_{(k)}$ for each value of the index $A$ and the solution is $1/2$-BPS.\par
As we have 4 chiral spinors in an $\mathcal{N}=2$ theory we can identify the soliton solution to be $1/2$ BPS in $\mathcal{N}=2$. With respect to the $\mathcal{N}=8$ theory, the solution is $1/8$ BPS.

\subsection{Supersymmetric solutions with fixed fluxes}
We have supersymmetric solitons when the bulk solution for given boundary conditions satisfies \,$Q_{1}=-\frac{1}{\sqrt{3}}\,Q_{2}$\,. For fixed flux boundary conditions, this requires
\begin{equation}
\psi_{1}=\frac{\psi_{2}\,x_{0}^{-2}}{\sqrt{3}}\:.
\end{equation}
Using \eqref{Pol} to determine $x_0$, we find that there are supersymmetric solitons for
\begin{alignat}{3}
&\psi_{1}=\sqrt{2}-\sqrt{3}\,\psi_{2}\,,
\qquad
&& x_{0}^{2}=\frac{\psi_{2}}{\sqrt{6}-3\,\psi_{2}}\,,
\qquad
&& 0<\psi_{2}<\frac{\sqrt{6}}{3}\,,
\label{bound}
\\[\jot]
& \psi_{1} =-\sqrt{2}-\sqrt{3}\,\psi_{2}\,,
\qquad
&& x_{0}^{2}=-\frac{\psi_{2}}{\sqrt{6}+3\,\psi_{2}}\,,
\qquad
&& -\frac{\sqrt{6}}{3}<\psi_{2}<0\,.
\end{alignat}
This reduces to the previous solution in \cite{Anabalon:2021tua} when $\psi_2 = \pm \frac{\sqrt{6}}{4}$.

For the fixed flux boundary conditions, we can also look for supersymmetric domain wall/Poincar\'{e}-AdS solutions.
As far as Poincar\'{e}-AdS solutions are concerned, we can start from the general expression of the Killing spinors in the appropriate parametrization of AdS$_4$, given in \cite{Anabalon:2021tua}. Considering Poincar\'{e}-AdS solutions in which $\varphi$ is compact, half of these spinors are not consistently defined and thus the solution is $1/2$-BPS in the $\mathcal{N}=8$ theory. In the absence of fluxes the Killing spinors can only obey  periodic boundary conditions along the $\varphi$-circle. We can switch on constant values of the four STU vector fields along the compact direction, defining Wilson-lines. This will not alter the local AdS geometry since the field strengths $\bar{F}^i_{\mu\nu}$ are all kept zero.  If we denote by $\epsilon_{I_\alpha},\,\epsilon^{I_\alpha}$, ($\alpha=1,\dots,4$,\, $I_1=1,2$,\, $I_2=3,4$,\, $I_3=5,6$,\, $I_4=7,8$) the $\mathcal{N}=8$ supersymmetry parameters, we choose those describing the $\mathcal{N}=2$ supersymmetry of the STU truncation to be the first two $I_1=A=1,2$. Restricting to the dilatonic field $\phi$ and setting $F^\Lambda_{\mu\nu}=0$, the supersymmetry variation of the $\mathcal{N}=8$ fermions read:
\begin{equation}
\begin{split}
\delta \psi^A_\mu &=\partial_\mu\epsilon^A+\frac{1}{4}\,\omega_\mu^{ab}\,\gamma_{ab}\,\epsilon^A-\frac{1}{2\sqrt{2}L}\,(A^1_\mu+\sqrt{3}\,A^2_\mu)\,\epsilon^{AB}\epsilon^B+\frac{\mathcal{W}}{2}\gamma_\mu\,\epsilon_A\,,
\\[1.5ex]
\delta \psi^{I_\alpha}_\mu &=\partial_\mu\epsilon^{I_\alpha}+\frac{1}{4}\,\omega_\mu^{ab}\,\gamma_{ab}\,\epsilon^{I_\alpha}-\frac{1}{2\sqrt{2}L}\,\left(A^1_\mu-\frac{1}{\sqrt{3}}\,A^2_\mu\right)\,\epsilon^{I_\alpha J_\alpha}\epsilon^{J_\alpha}+\frac{\mathcal{W}}{2}\gamma_\mu\,\epsilon_{I_\alpha}\,,\quad\;(\alpha\neq1)\,.   \end{split}
\end{equation}
In the Poincar\'{e}-AdS solution \,$\phi=0$\, and \,$\mathcal{W}=1/L$\,.
From the minimal couplings to the vector fields, we can infer the dependence of the Killing spinors on the fluxes:
\begin{align}
 \epsilon^A(\varphi,r)=e^{i\,\omega_1\varphi}\,\mathring{ \epsilon}^A(r)\,,
\qquad\qquad
 \epsilon^{I_\alpha}(\varphi,r)=e^{i\,\omega_2\varphi}\,\mathring{ \epsilon}^{I_\alpha}(r)\qquad(\alpha\neq 1)\,,\quad
 \label{epsiomega}
\end{align}
where:
\begin{equation}
 \omega_1=\frac{\pi}{\sqrt{2}\,\Delta}\,(\psi_1+\sqrt{3}\,\psi_2)\,,
\qquad\quad \omega_2=\frac{\pi}{\sqrt{2}\,\Delta}\,\left(\psi_1-\frac{1}{\sqrt{3}}\,\psi_2\right)\,.
\label{omegas}
\end{equation}
By appropriately choosing the fluxes, we can have the Killing spinors satisfy both periodic and anti-periodic boundary conditions. One needs to restrict $\psi_1,\psi_2$ to satisfy:
\begin{equation}
\psi_1=\frac{3\,m+n}{2\sqrt{2}}\,,\qquad\quad
\psi_2=\frac{1}{2}\,\sqrt{\frac{3}{2}}\,(n-m)\,,
\label{psimn}
\end{equation}
where $m,n$ are even or odd integers for periodic or anti-periodic boundary conditions, respectively.

Let us now consider domain wall solutions obtained by setting $Q_1=Q_2=0$ in the solitonic ones. The Killing spinors are obtained from eqs.\ \eqref{KS2} by setting $\alpha_-(x)=0$, being $f(x)=1$.
The latter feature implies that the $\varphi$-circle, as opposed to the charged solition case, is non-contractible. In the absence of fluxes, $\omega=0$, as pointed out in the previous section, and for each $A$ we have two linearly independent solutions $\epsilon^A_{(k)}$ so that the solution preserves therefore $1/2$ of the $\mathcal{N}=2$ supersymmetries of the model and all Killing spinors can only obey periodic boundary  conditions on the $S^1$ spanned by $\varphi$. When embedding the solution in the $\mathcal{N}=8$ model, one can verify that Killing spinor equations for $\epsilon^{I_\alpha}$ ($\alpha\neq 1$) are the same as those in the spinors $\epsilon^{A}$ and thus admit the same solutions. The domain wall solutions, in the absence of fluxes, are therefore $1/2$-BPS in the $\mathcal{N}=8$ model. Switching the fluxes on, as constant potentials along $S^1$, the supersymmetry counting does not change and we still have a $1/2$-BPS solution in the $\mathcal{N}=2$ model.  Within the $\mathcal{N}=8$ model, $\epsilon^{I_\alpha}$ ($\alpha\neq 1$) satisfy the same Killing spinor equation as $\epsilon^{A}$, though with a different combination of fluxes, which implies that the Killing spinors will acquire a dependence on $\varphi$ given in \eqref{epsiomega}, with $\omega_1,\,\omega_2$ given in \eqref{omegas}. Thus choosing the fluxes to satisfy \eqref{psimn}, for suitable choices of the integers $m,\,n$, we can have the Killing spinors satisfy either periodic or anti-periodic boundary conditions, which are both allowed on the domain wall backgrounds.

\subsection{Supersymmetric solutions with fixed charges}
We get supersymmetric solutions if
\begin{equation}
q_{1}=-\frac{q_{2}}{\sqrt{3}}\:.
\end{equation}
We find that for every value $\left\vert q_{2}\right\vert <\sqrt{6}/16$ there are
two charged solitons, located at the two $x_\textsc{s}$ that follow from the
equation:
\begin{equation}
q_{2}=\pm\sqrt{6}\:\frac{\xs^{3}}{\left(3\,\xs^{2}+1\right)^{2}}\:,
\end{equation}
the signs depending on whether the charge is positive or negative. One of the
solitons is located at $\xs>1$ and the other is located at $\xs<1$. As previously noted, in addition to these supersymmetric solutions this choice of boundary conditions has non-supersymmetric solutions with vanishing scalar.

\section{Phase structure}
\label{phase}

\subsection{Euclidean action}
To compute the Euclidean action we continue the metric to a real Euclidean metric $g_{\textsc{e}}$ with Euclidean time $\tau \in [0,\beta]$, where $\beta$ is the inverse of the temperature. The magnetic gauge field and the dilaton remain invariant under this change yielding a periodic bosonic solution in $\beta$.

We remark that the thermal partition function is not compatible with supersymmetry unless $\beta=\infty$, as the thermal partition function requires the fermions to be antiperiodic in Euclidean time. Hence, the supersymmetric solutions exist only at zero temperature. However, the non-supersymmetric bosonic solutions are well defined for all $\beta$.

The Euclidean action $S_{\textsc{e}}$ has several contributions:
\begin{equation}
\frac{S_{\textsc{e}}}{V}=I_{\text{bulk}}+I_{\textsc{gh}}+I_{\textsc{bk}}+I_{\text{ct}}+I_{\phi}\:,
\end{equation}
where
\begin{equation}
V=\beta\,\Delta\,\Delta_z\,,
\qquad\quad
\Delta_{z}=\int dz\,.
\end{equation}
The $I_{\text{bulk}}$ term is the bulk contribution,
\begin{equation}
I_{\text{bulk}}=\lim_{\epsilon\to 1^{-}}\;\frac{1}{\kappa}\:\int\limits_{x_{0}}^{\epsilon}dx\;\sqrt{g_{\textsc{e}}}\,\left(  -\frac{R}{2}+\frac{1}{2}\left(\partial\phi\right)^{2} -\frac{3}{L^{2}}\,\cosh\left(\sqrt{\frac{2}{3}}\,\phi\right) -\frac{1}{4}\,e^{3\sqrt{\frac{2}{3}}\,\phi}\left(F^{1}\right)^{2} -\frac{1}{4}\,e^{-\sqrt{\frac{2}{3}}\,\phi}\left(
F^{2}\right)^{2}\right),
\end{equation}
while $I_{\textsc{gh}}$ is the Gibbons-Hawking term, $I_{\textsc{bk}}$ the Balasubramanian-Krauss counterterm, $I_{\text{ct}}$ is a divergent counterterm for the scalar field and $I_{\phi}$ is a finite counterterm that ensures that the action principle is well posed for the scalar field:
\begin{equation}
\begin{split}
I_{\textsc{gh}}&=-\frac{1}{\kappa}\;\lim_{\epsilon\to1^{-}} K\,\sqrt{h}\,,
%\qquad\quad
\\[1.5ex]
I_{\textsc{bk}}&=\frac{2}{\kappa\,L}\lim_{\epsilon\to1^{-}}\sqrt{h}\,,
\\[1.5ex]
I_{\text{ct}}&=\frac{1}{2\,\kappa\,L}\lim_{\epsilon\to1^{-}}\sqrt{h}\,\phi^{2}\,,
%\qquad\quad
\\[1.5ex]
I_{\phi}&=-\frac{L^{2}}{\kappa}\:\frac{1}{3\sqrt{6}}\:\phi_{0}^{3}\,.
\end{split}
\end{equation}
It is straightforward to find that
\begin{equation}
\begin{split}
I_{\text{bulk}} &  =\frac{1}{\kappa}\,\lim_{\rho\to\infty}\left(\frac
{\rho^{3}}{L^{4}}+\frac{3\,\rho}{8\,\eta^{2}}-\frac{\mu}{L^{2}}+\frac{L^{2}}{2\,\eta^{3}}\right),
\\[2ex]
I_{\textsc{gh}} &=-\frac{1}{\kappa}\,\lim_{\rho\to\infty}\left(\frac{3\,\rho^{3}}{L^{4}}+\frac{9\,\rho}{8\,\eta^{2}}-\frac{3\,\mu}{2\,L^{2}}
+\frac{3\,L^{2}}{2\,\eta^{3}}\right),
\\[2ex]
I_{\textsc{bk}} & =\frac{1}{\kappa}\,\lim_{\rho\to\infty}\left(\frac{2\,\rho^{3}}{L^{4}}-\frac{\mu}{L^{2}}\right),
\\[2ex]
I_{\text{ct}}  &=\frac{1}{\kappa}\,\lim_{\rho\to\infty}\left(\frac{3\,\rho}{4\,\eta^{2}}+\frac{3\,L^{2}}{4\,\eta^{3}}\right),
\\[2ex]
I_{\phi} &=\frac{L^{2}}{\kappa}\:\frac{1}{4\,\eta^{3}}\,,
\end{split}
\end{equation}
and we also have
\begin{equation}
\frac{S_{\textsc{e}}}{V}=-\frac{\mu}{2\,L^{2}\,\kappa}\:.
\end{equation}
We found that the result is the same for both solutions with the definition of $\mu$ given in \eqref{mu}. These configurations do not have an associated entropy, so we get that their free energy is just their energy.

\subsection{Fixed fluxes $\psi_1$ and $\psi_2$}
\begin{figure}[!h]
\centering
\includegraphics[scale=0.65]{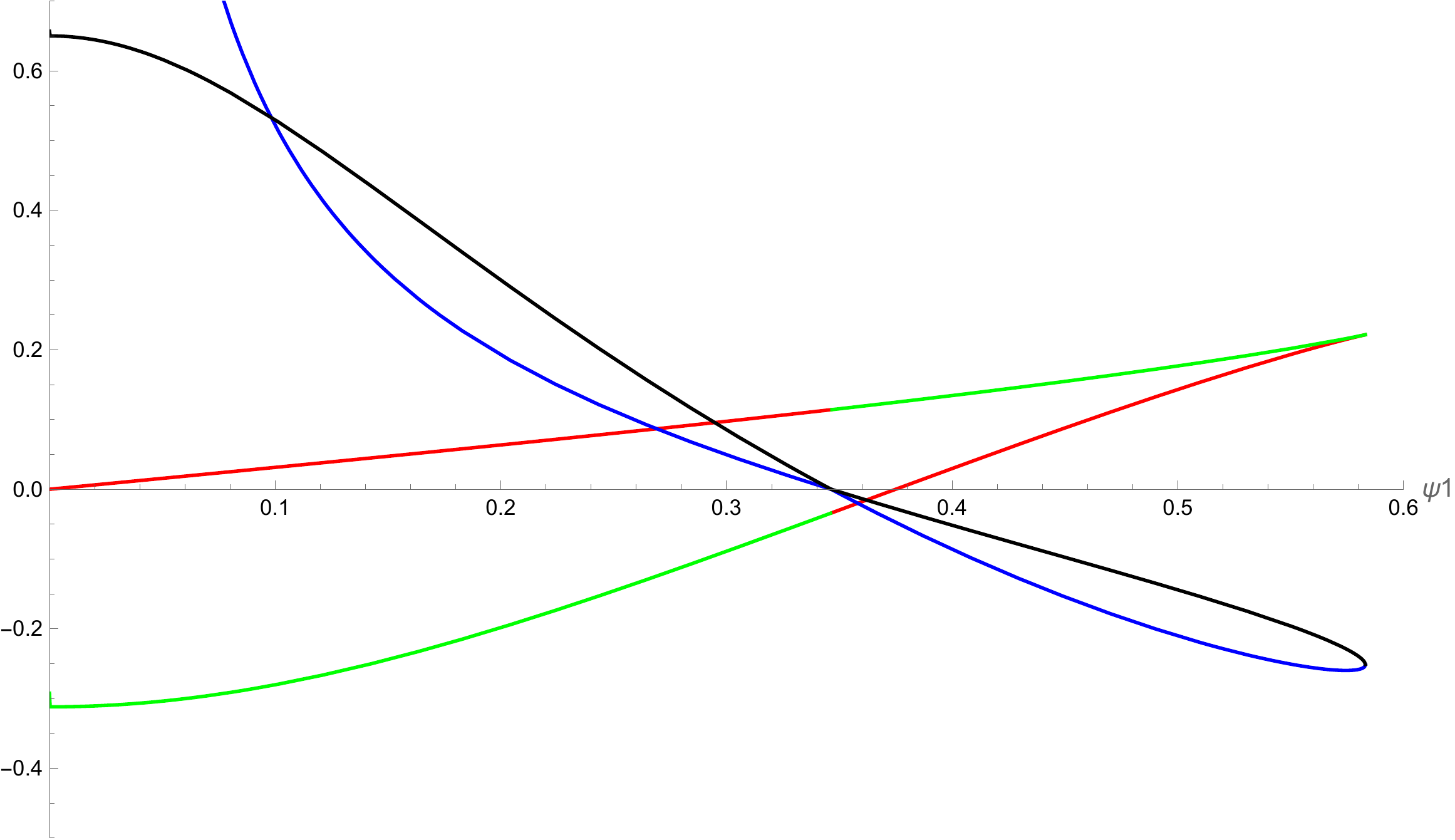}
\caption{Rescaled free energy $\frac{G_{\phi}}{\left\vert G_{0}\right\vert }$ (red, green) and rescaled vev $\left\langle \mathcal{O}\right\rangle \Delta/\pi$ (black, blue) for $\psi_{2}=\pm 0.6$. Different
colours are used to represent different branches of the solution.}
\label{fig:fig3}
\end{figure}

\begin{figure}[!h]
\centering
\includegraphics[scale=0.7]{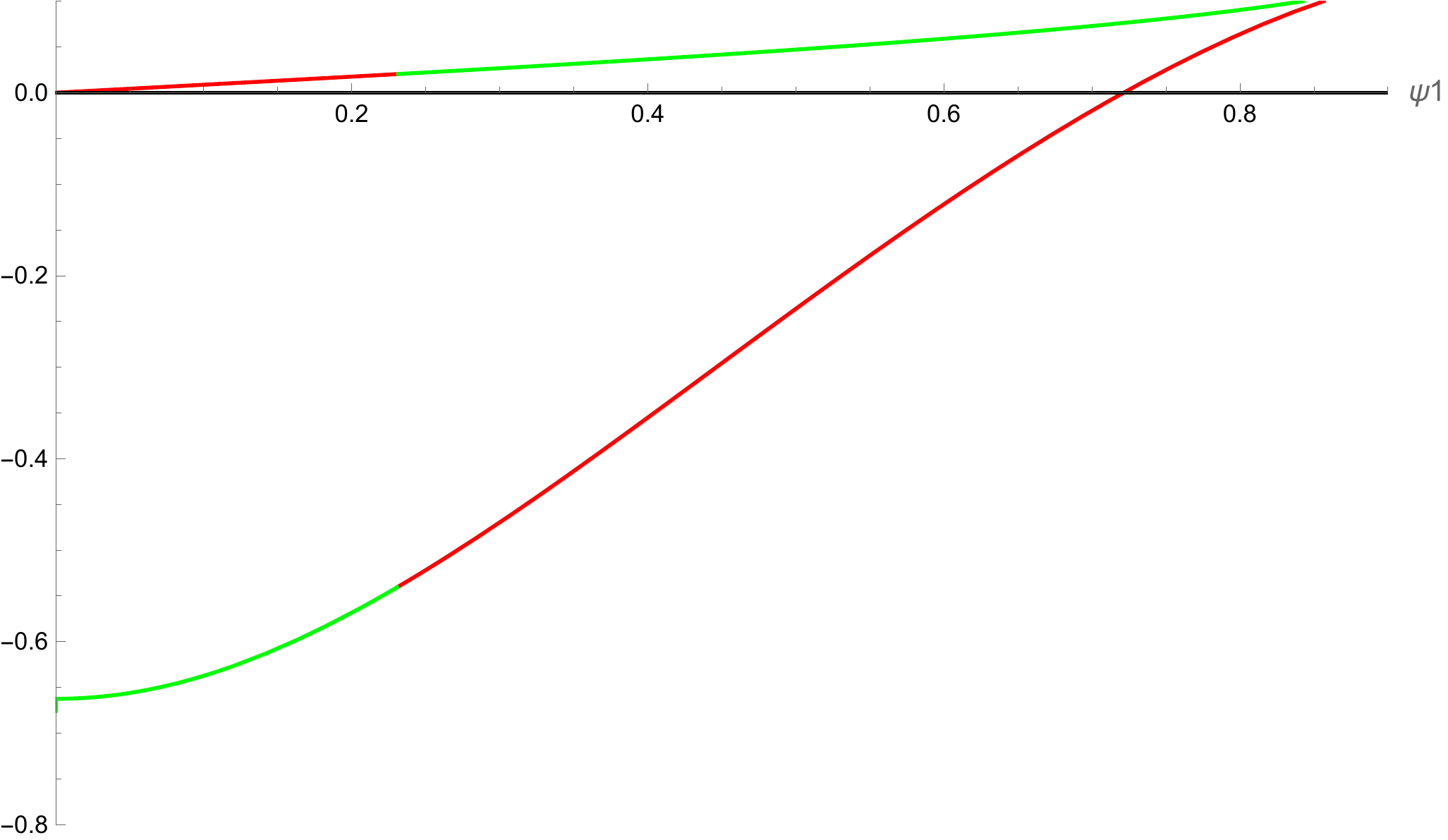}
\caption{Rescaled free energy $\frac{G_{\phi}}{\left\vert G_{0}\right\vert }$ at $\psi_{2}=\pm 0.4$. The different colours represent
different branches of the solution.}
\label{fig:fig4}%
\end{figure}

\begin{figure}[!h]
\centering
\includegraphics[scale=0.7]{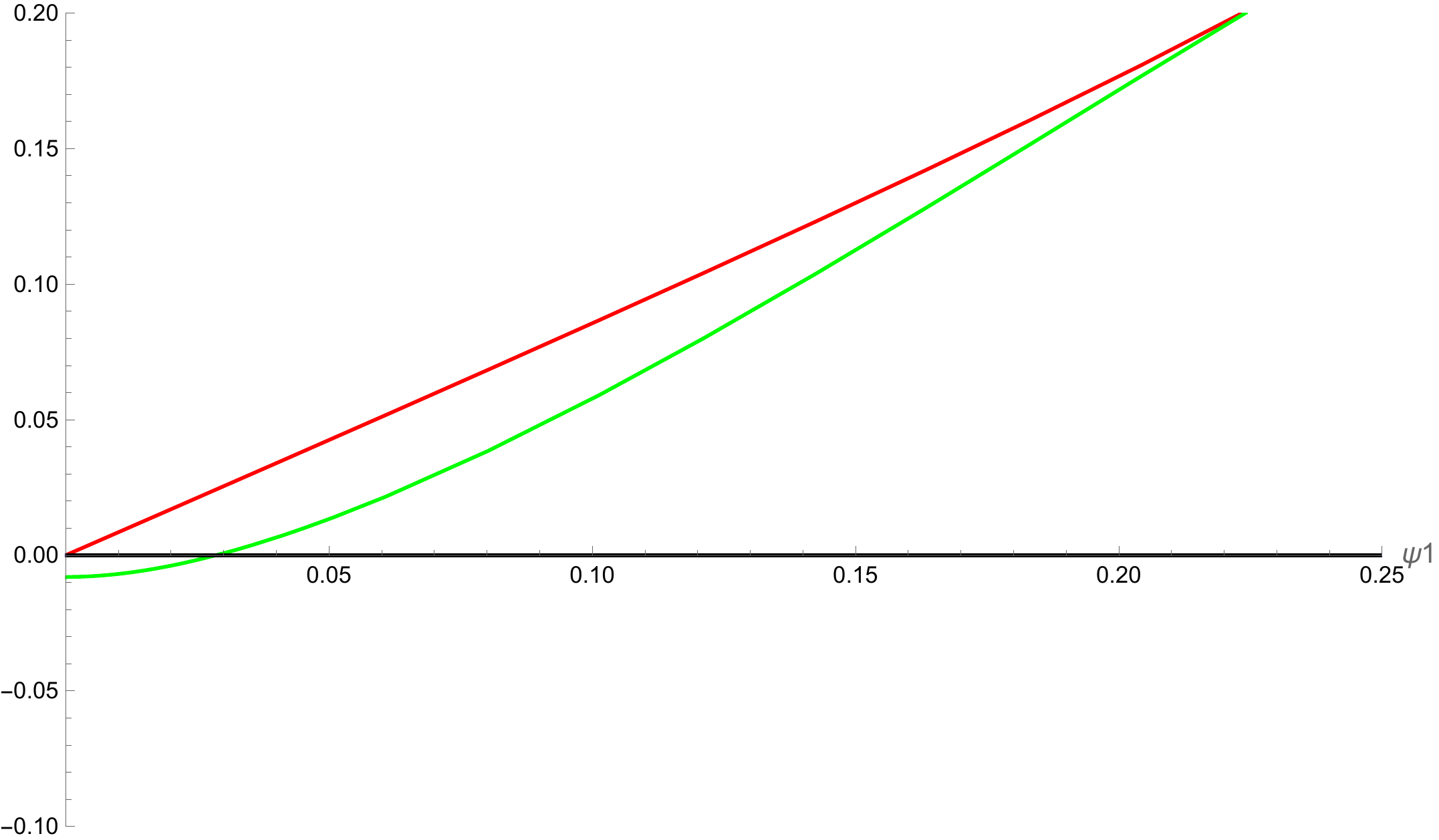}
\caption{Rescaled free energy $\frac{G_{\phi}}{\left\vert G_{0}\right\vert}$ for $\psi_{2}=\pm 0.8$. The different
colours represent different branches of the solution. We note that, for these
values of $\psi_{2}$, only the $x>1$ solutions are necessary to describe the
phase diagram.}
\label{fig:fig5}%
\end{figure}

The free energy density of the hairy soliton solution is
\begin{equation}
\begin{split}
\frac{S_{\textsc{e}}}{V}  &  =\frac{G_{\phi}}{\Delta\,\Delta_{z}} =\frac{M}{\Delta\,\Delta_{z}}=
%\\[1ex]&=
-\frac{\mu}{2\,\kappa\,L^2}
=\pm\frac{2}{3\,\eta\,\kappa}\left(  3\,Q_{1}^{2}-Q_{2}^{2}\right) =
\\[1ex]
& =\pm\frac{8\pi^{3}L^{2}}{3\,\Delta^{3}\,\kappa}\:\frac{x_{0}\left\vert 2\,x_{0}^{2}\,\psi_{1}^{2}+\psi_{1}^{2}-\psi_{2}^{2}\right\vert \left(  3\,\psi_{1}^{2}\,x_{0}^{4}-\psi_{2}^{2}\right)}{\left(x_{0}^{2}-1\right)^{2}}\:.
\end{split}
\end{equation}
The free energy of the AdS soliton is given by \cite{Horowitz:1998ha}
\begin{equation}
G_{0}=-\frac{32}{27}\:\frac{\pi^{3}L^{2}}{\Delta^{3}\,\kappa}\:\Delta\,\Delta_{z}\:.
\end{equation}
We can use this energy as a convenient normalization%
\footnote{%
As previously remarked, the AdS soliton does not satisfy the boundary conditions except at $\psi_{1}=\psi_{2}=0$.}
and plot the ratio
\begin{equation}
\frac{G_{\phi}}{\left\vert G_{0}\right\vert }=\pm\frac{9}{4}\:\frac{x_{0}\left\vert 2\,x_{0}^{2}\,\psi_{1}^{2}+\psi_{1}^{2}-\psi_{2}^{2}\right\vert
\left(3\,\psi_{1}^{2}\,x_{0}^{4}-\psi_{2}^{2}\right)}{\left(x_{0}^{2}-1\right)^{2}}\:,
\end{equation}
together with the value of the rescaled vev $\left\langle \mathcal{O}\right\rangle \Delta$ in Figures \ref{fig:fig3}, \ref{fig:fig4} and \ref{fig:fig5}.
In these plots, the red line is the solution associated with the root
$x_{0,1}$ and the green line is the solution associated with the root $x_{0,3}$ (defined in Appendix \ref{app:roots}).
For the rescaled vev
$\left\langle \mathcal{O}\right\rangle \Delta$, the blue line is the
vev of the solution $x_{0,1}$ and the black line is the vev of the
solution $x_{0,3}$. The region with negative vev are the
solutions at $x>1$ and the region with positive vev are the solutions at $x<1$. The susy soliton is located at the intersection of the free energy with the $\psi_{1}$ axis.

When $\psi_{1}=0$, the family with lower energy goes to a state with negative energy and finite vev, which is the soliton with only $\psi_{2}\neq
0$. The family with higher energy goes to a state with zero energy and infinite
$\left\langle \mathcal{O}\right\rangle$.
We also found that, if we set
$\Delta=\delta\psi_{1}^{-4}$, the metric goes to the domain wall solution with
$f(x)=1$ and finite $\psi_{2}$.

\medskip

\subsection{Fixed $q_\Lambda$}
Now we consider the Legendre transform of the Euclidean action %
\begin{equation}
\begin{split}
\frac{F_{\phi}}{V}&=\frac{S_{\textsc{e}}}{V}-\left.\left\langle J_\Lambda^{\nu}\right\rangle A^\Lambda_{\nu}\right|_{x\to1}=
\\[1ex]
&=-\frac{\mu}{2\,L^{2}\,\kappa}
-\langle J_1^{\phi}\rangle\,
Q_1\left(1-x_0^{-2}\right)
-\langle J_2^{\phi}\rangle\, Q_2\left(1-x_0^2\right)=
\\[1ex]
&=\pm\frac{2}{3\,\eta\,\kappa}\left(3\,Q_{1}^{2}-Q_{2}^{2}\right)\pm\frac{2\,Q_1^2}{\eta\,\kappa}\left(1-x_0^{-2}\right)\mp\frac{2\,Q_2^2}{\eta\,\kappa}\left(1-x_0^2\right)\,,
\end{split}
\end{equation}
where the minus sign is for the solutions with $x<1$ and the plus sign is for the solutions with $x>1$.

\paragraph{Supersymmetric solutions.}
As described in Section \ref{susy}, we have supersymmetric solitons for $q_2 = -\sqrt{3}\,q_1$, and the solutions of \cite{Anabalon:2021tua} with $\phi =0$ also satisfy these boundary conditions. It is therefore particularly interesting to plot the phase diagram in this sector of the fixed $q_\Lambda$ boundary conditions. We will see that it leads to a significant surprise.

The parameters of \cite{Anabalon:2021tua} are related to the parameters here by
\begin{equation}
q_1 =\frac{\Delta^{2}}{4\pi^{2}L}\:\frac{Q}{\sqrt{8}\,L^2}\:. \quad
\end{equation}
The free energy for the Einstein-Maxwell solutions is \cite{Anabalon:2021tua}%
\footnote{%
Here we fix a wrong sign in front of $Q^2$ in equations (44) and (45) of \cite{Anabalon:2021tua}.
}
\begin{equation}
\begin{split}
\frac{F_{\textsc{em}}}{\Delta\,\Delta_z} =-\frac{m}{2\,\kappa\,L^{2}}+\frac{2\,Q^2}{\kappa\,L^2\,r_0} =\frac{2\pi^3L^2}{\Delta^3\,\kappa}\:X^2\,(5-4\,X)\:,
\end{split}
\end{equation}
with $q_1^2=2^{-7}\,X^3\,(4-3\,X)$ and $X=\frac{r_0\,\Delta}{\pi L^2}$\,.
For the supersymmetric hairy solutions we find
\begin{equation}
\frac{F_{\phi}}{|G_0|}=\frac{27}{\sqrt{2}}\,|q_1|\:.
\end{equation}
Both the Einstein-Maxwell and the hairy solitons exist for $q_1^2 \leq 2^{-7}$. At this point all the different branches of solutions merge yielding a unique supersymmetric soliton there.

\begin{figure}[H]
\centering
\includegraphics[scale=0.6]{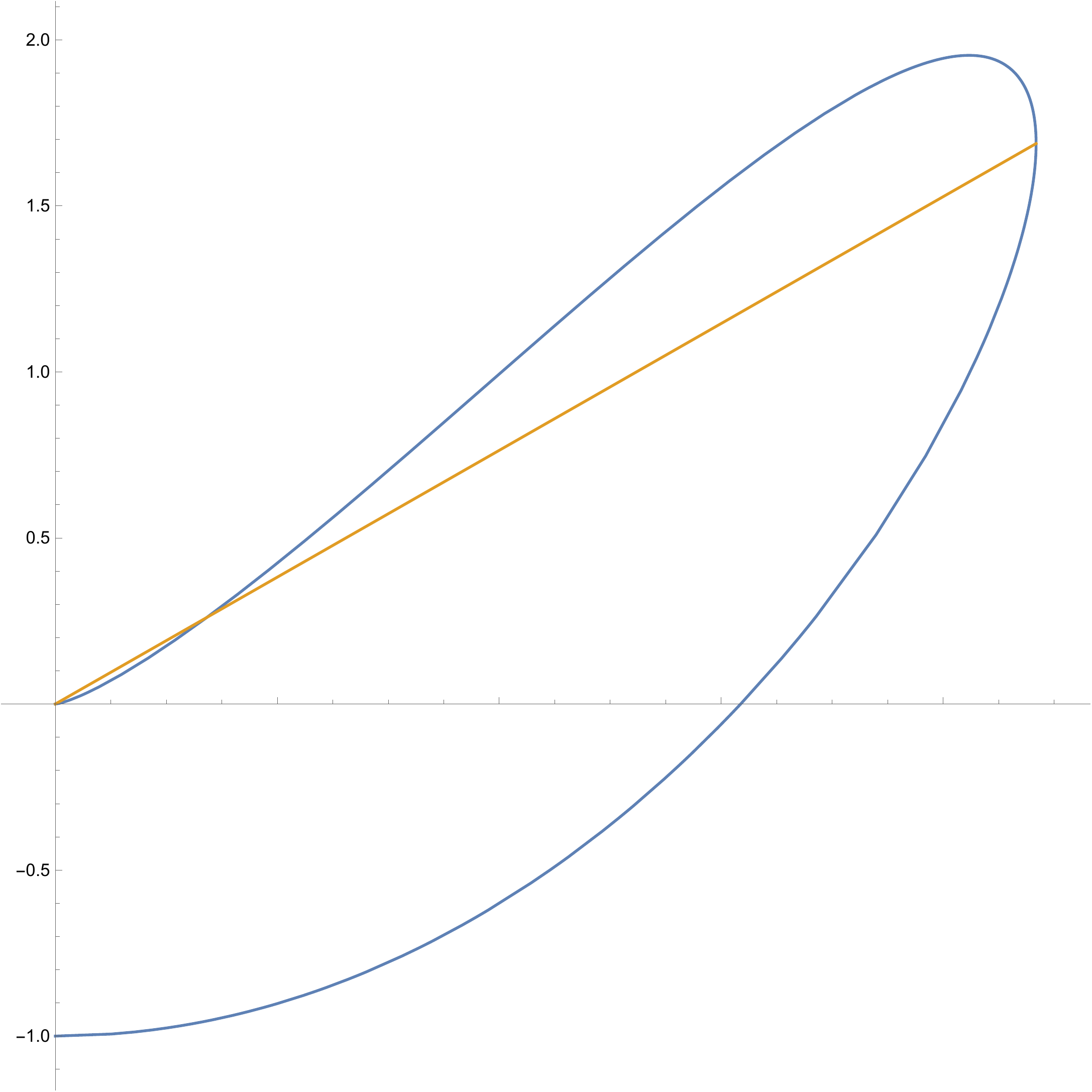}
\caption{Rescaled free energy $\frac{F}{\left\vert G_{0}\right\vert}$ vs $q_1$ on the supersymmetric shell $q_2=-\sqrt{3}\,q_1$. The yellow line represents the hairy supersymmetric solitons. Note that there are two distinct solitons for each point on this curve that coalesce at the right-hand end. The non-supersymmetric pure Einstein-Maxwell solitons are shown in blue. It is surprising that there are non-supersymmetric solutions with lower free energy (and lower energy), for the same boundary conditions and asymptotic charges, than a supersymmetric solution. This issue discussed in Section \ref{PET}.}
\label{fig:fig6}%
\end{figure}

In Figure \ref{fig:fig6}  we plot $\frac{F}{|G_0|}$ as a function of $q_1$ for $q_2 = - \sqrt{3}\,q_1$. Surprisingly, we see that one branch of the non-supersymmetric Einstein-Maxwell solutions has lower free energy than the supersymmetric hairy solution (they also have lower energy). At first sight this seems surprising as we would expect the supersymmetric solutions to saturate a BPS bound. As anticipated in the Introduction, this result is not in contradiction with the positive energy theorem if we include among the boundary conditions defining a solution also those applying to the asymptotic Killing spinor, which is the central ingredient in the proof of the theorem. We shall expand on this point in the next subsection.\par\smallskip

\begin{figure}[H]
\begin{subfigure}{.5\textwidth}
\centering
\includegraphics[scale=0.6]{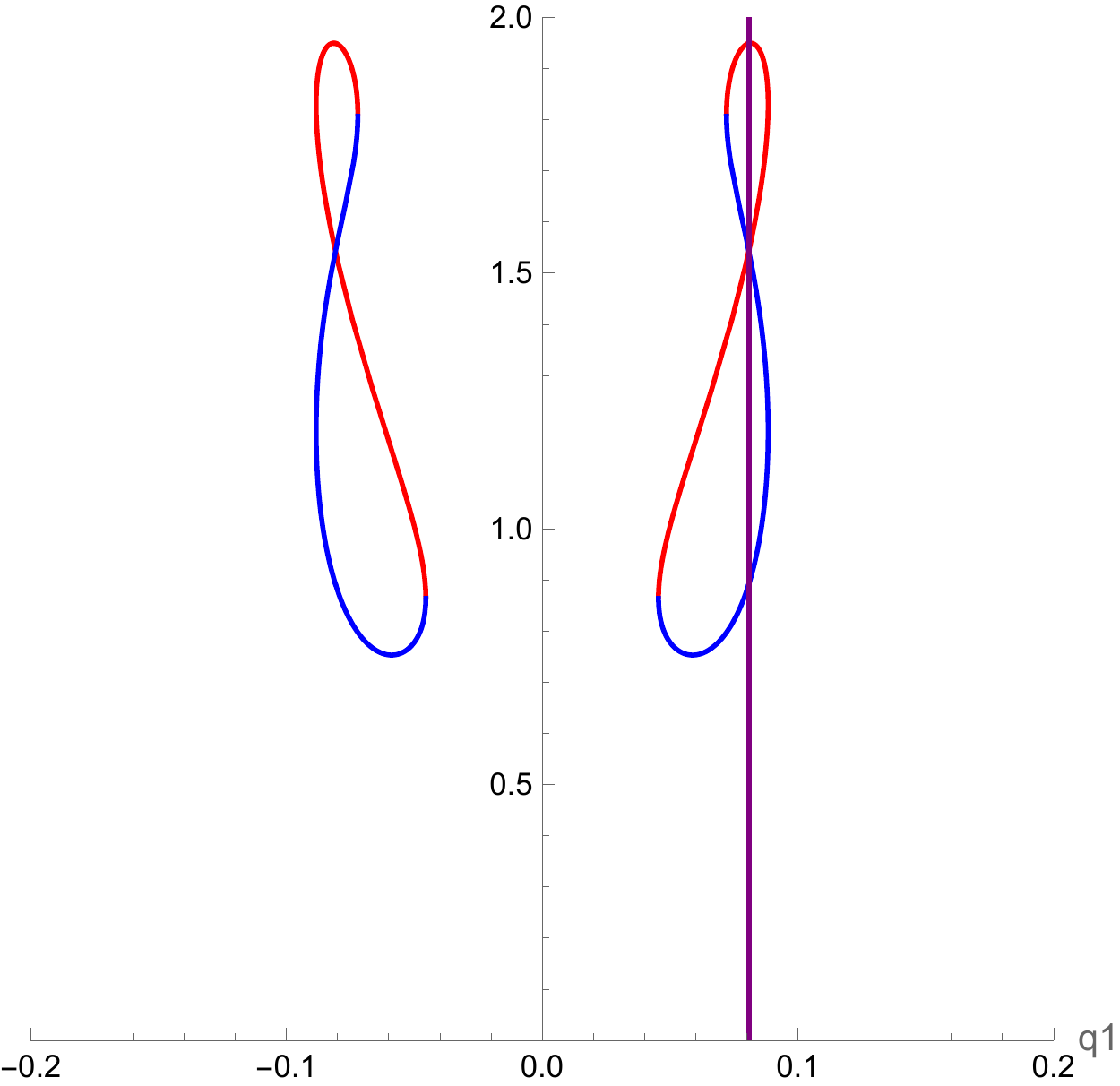}
\caption{$q_2=-0.14$}
\label{subfig:sfig1}
\end{subfigure}
\begin{subfigure}{.5\textwidth}
\centering
\includegraphics[scale=0.6]{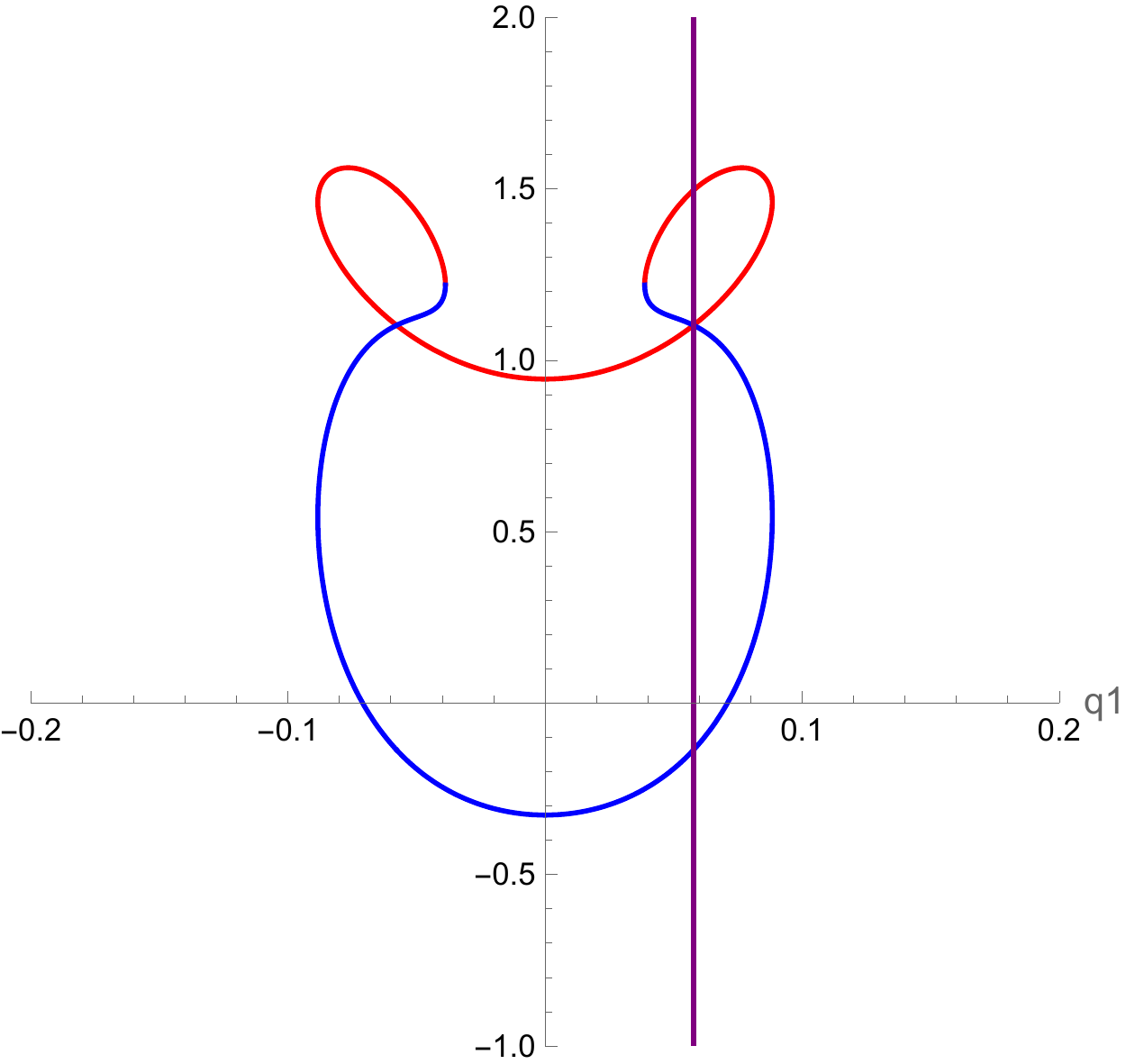}
\caption{$q_2=-0.1$}
\label{subfig:sfig2}
\end{subfigure}
\begin{subfigure}{.5\textwidth}
\centering
\includegraphics[scale=0.6]{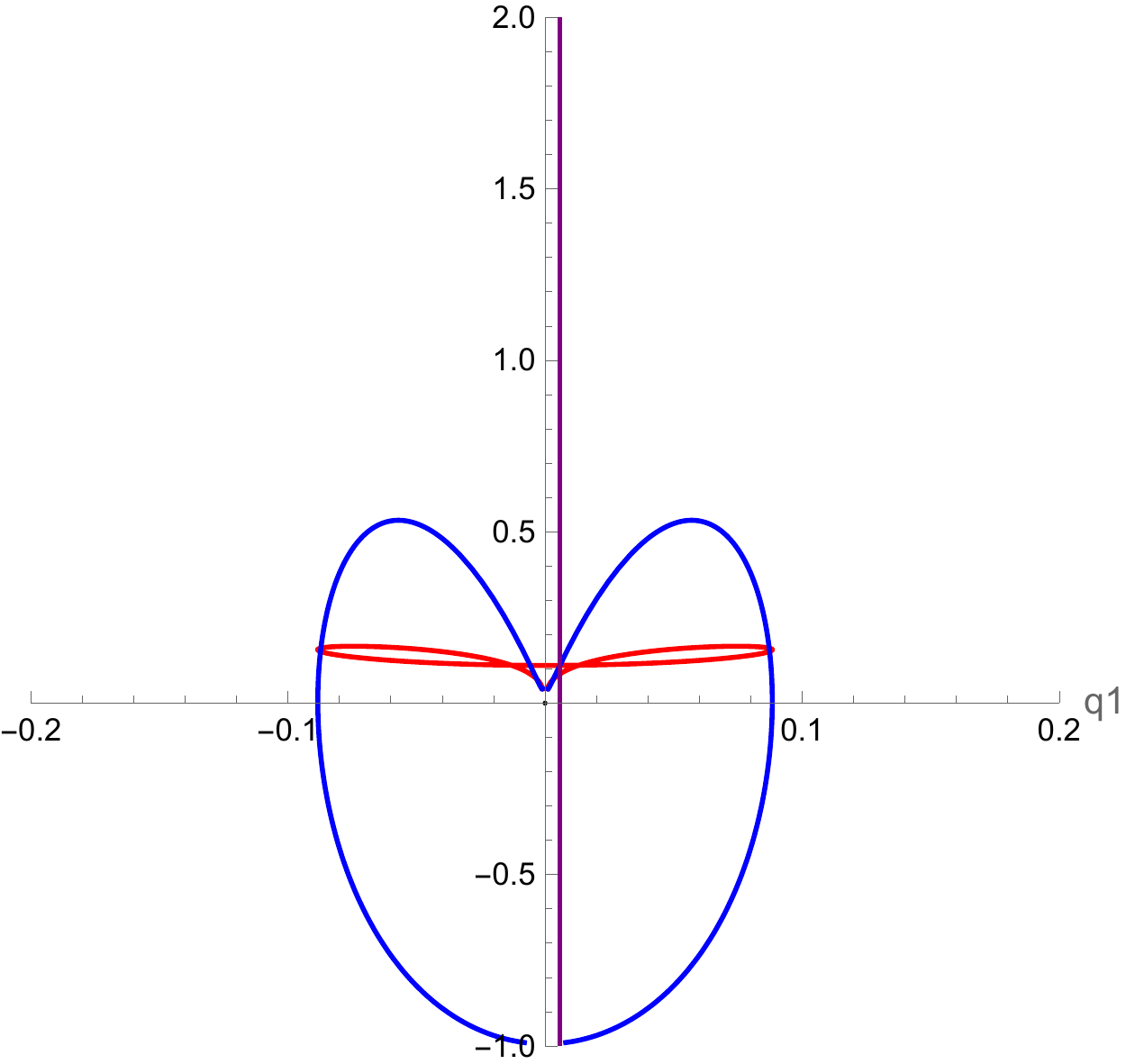}
\caption{$q_2=-0.01$}
\label{subfig:sfig3}
\end{subfigure}
\begin{subfigure}{.5\textwidth}
\centering
\includegraphics[scale=0.6]{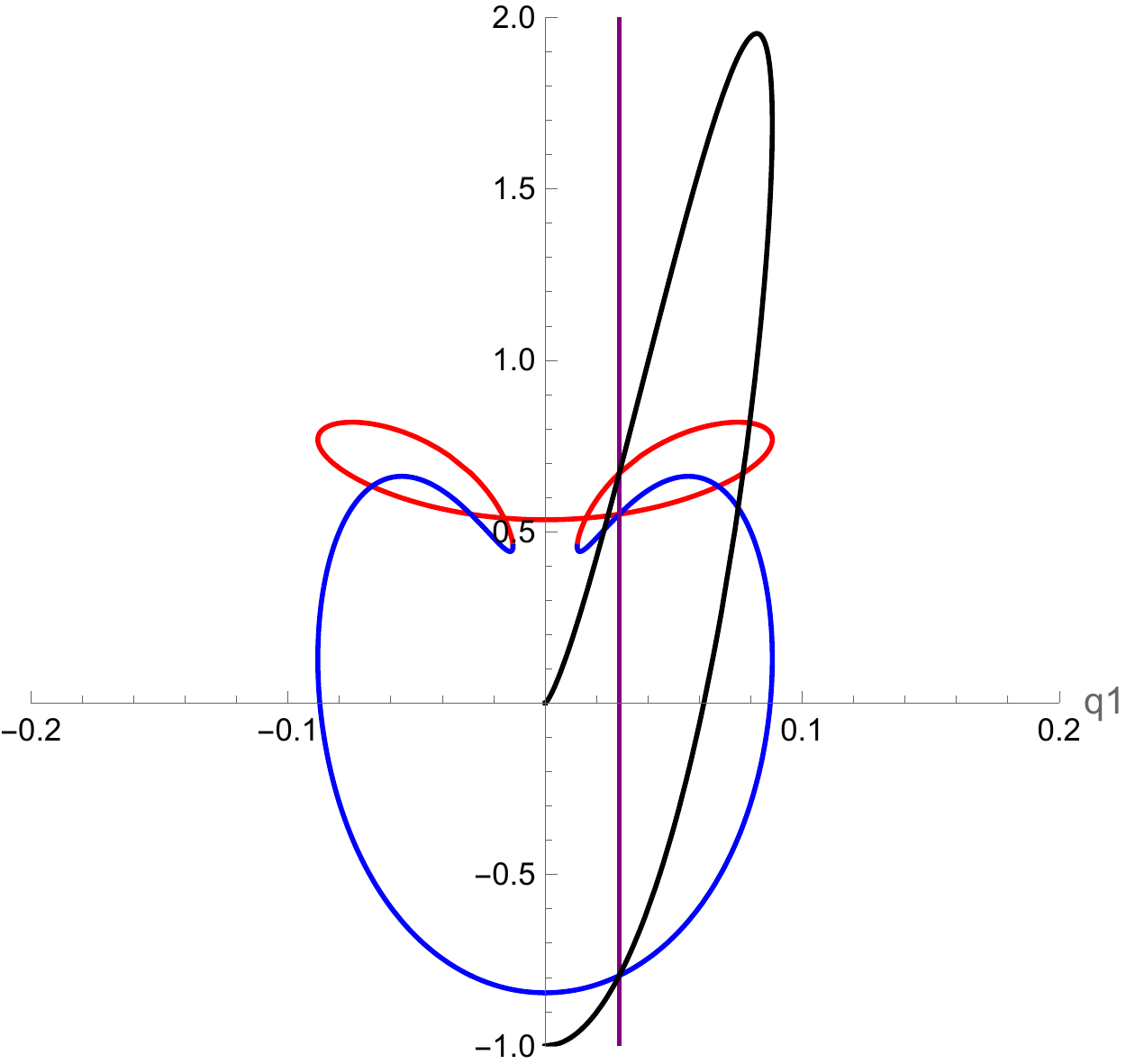}
\caption{$q_2=-0.05$}
\label{subfig:sfig4}
\end{subfigure}
\caption{Rescaled free energy $\frac{F_{\phi}}{\left\vert G_{0}\right\vert }$ vs.\ $q_1$ for different values of $q_2$.
The supersymmetric solutions are located at the intersection of the blue and red branches with the purple line. The other intersections of the purple line with the red and blue branches are solutions of the pure Einstein-Maxwell system at zero dilaton. This is verified in the last panel by plotting the Einstein-Maxwell solutions in black and observing how it intersects the blue and red branches at exactly the point where they also intersect the purple line. We remark that the Einstein-Maxwell soliton represented by the black line is not part of the last phase diagram except at the points of intersection with the purple vertical line; otherwise it does not satisfies the same boundary conditions of the hairy configurations.}%
\label{fig:fig7}
\end{figure}

In Figure \ref{fig:fig7} we plot the phase diagram as a function of $q_1$ for fixed $q_2$ for several values of $q_2$. Is possible to see that, for a given value of the charges, there are from 0 to 4 solitons, as we discussed earlier. The purple line indicates the value of $q_1$ that satisfies the supersymmetric condition $q_2=-\sqrt{3}\,q_1$ for the given value of $q_2$.
We note that the supersymmetric solutions are located at the intersection of this line with the blue and red branches. The solutions on the purple line which are above or below this intersection do not exist as hairy solutions.
However, there is a non-supersymmetric solution satisfying the same boundary conditions with zero dilaton, that can be found at the spot where the hairy solutions does not exist. To show this, we plot the free energy of the Einstein-Maxwell solitons in black in the \subref{subfig:sfig4} panel. It is possible to see that, when the blue and the red solutions do not exist, the black solution takes their place in the free energy diagram. This is what is expected from Figure \ref{fig:fig6}. We remark that the Einstein-Maxwell soliton represented by the black line only satisfies the boundary conditions when $q_2 = -\sqrt{3}\, q_1$.

\subsubsection{Boundary conditions and the positive energy theorem.}\label{PET}
In order to compare the energy of the non-supersymmetric solutions of \cite{Anabalon:2021tua}, for fixed charges at infinity, with that of the supersymmetric hairy ones, in light of the positive energy theorem, the former should admit a spinor field which asymptotes a Killing spinor of the latter at radial infinity, namely satisfy the Killing spinor equation with the same boundary conditions, up to terms of order $1/r^2$. This asymptotic Killing spinor should then satisfy anti-periodic boundary conditions along the $S^1$ at the radial boundary.
This amounts, on the non-supersymmetric solution, to the requirement:
\begin{equation}
\frac{1}{2\sqrt{2}\,L}\lim_{r\rightarrow \infty}\left(A^1_\varphi+\sqrt{3}\,A^2_\varphi\right)\Delta=\pi\,n\,,
\quad\;
n\in \mathbb{Z}\,,
\quad
n\,\,\text{odd}\,,
\label{cond1}
\end{equation}
which implies
\begin{equation}
q_1^2=2^{-7}\,n^2\,X^2\,,
\qquad\quad
X\equiv \frac{r_0\,\Delta}{\pi L^2}\,.
\end{equation}
Equating this expression with $q_1^2=2^{-7}\,X^3\,(4-3\,X)$, which is required by the regularity of the soliton, we find real solutions in $X$ only for $n=\pm 1$. These solutions are
\begin{equation}
X=1\,,\:\frac{1}{3}
\end{equation}
and correspond to
\begin{equation}
q_1^2=2^{-7}
\quad\;\text{or}\;\quad\,
q_1^2=\frac{2^{-7}}{9}\,,
\end{equation}
respectively.\par
The ratio of the free energies, in terms of $X$, reads:
\begin{equation}
\frac{F_{\textsc{em}}}{F_\phi}=\frac{(5-4\,X)\, X^2}{\sqrt{(4-3\,X)\,X^3}}\,.
\end{equation}
At $X=1$ the ratio is $1$ since the two solutions coincide. At $X=1/3$ the solution of \cite{Anabalon:2021tua} is non-supersymmetric and the above ratio is:
\begin{equation}
    \left.\frac{F_{\textsc{em}}}{F_\phi}\right\vert_{X=\frac{1}{3}}=\left.\frac{(5-4\,X) \,X^2}{\sqrt{(4-3\,X) \,X^3}}\right\vert_{X=\frac{1}{3}}=\,\frac{11}{9}\,>\,1\,,
\end{equation}
that is the energy of the non-supersymmetric solution is greater than that of the supersymmetric one with the same boundary conditions and asymptotic Killing spinors.

\section*{Acknowledgements}
SFR is supported in part by STFC through grant ST/T000708/1, and by a grant from the Simons Foundation, and his work was performed in part at the Aspen Center for Physics, which is supported by National Science Foundation grant PHY-1607611. The
research of AA is supported in part by the Fondecyt Grants 1210635, 1221504 and 1181047 and
by the FAPESP/ANID project 13231-7.

%---- APPENDICES ------------------------------------------

\newpage

\appendix
%\phantomsection
\addtocontents{toc}{\protect\setcounter{tocdepth}{1}}
\addtocontents{toc}{\protect\addvspace{3.5pt}}%
\numberwithin{equation}{section}%
\numberwithin{figure}{section}%

\section{Spinor conventions and SUSY for $\N=2$ models}\label{app:susy}
We shall use the Majorana basis for the Clifford algebra:
\begin{equation}
\gamma^{0}=-i
\begin{pmatrix}
0\! & \sigma_{2}\\
\sigma_{2} & 0
\end{pmatrix},
\quad\;
\gamma^{1}=-
\begin{pmatrix}
\!\sigma_{3}\! & 0\\
0 & \sigma_{3}%
\end{pmatrix},
\quad\;
\gamma^{2}=i
\begin{pmatrix}
0 & -\sigma_{2}\\
\sigma_{2} & 0
\end{pmatrix},
\quad\;
\gamma^{3}=
\begin{pmatrix}
\sigma_{1} & 0\\
0 & \sigma_{1}%
\end{pmatrix},
\end{equation}
in which it is possible to pick the charge conjugation matrix to be
\,$C=\gamma_{0}$. We also define the matrix $\gamma^5$ as:
\begin{equation}
\gamma^5=i\,\gamma^0\gamma^1\gamma^2\gamma^3\:,
\end{equation}
We shall use $\mathcal{N}=2$ chiral supersymmetry parameters $\epsilon^A,\,\epsilon_A$ \,($A=1,2$), defined as the chiral components of two Majorana spinors $\eM$:
\begin{equation}
\eM=\epsilon^A+\epsilon_A\,,
\end{equation}
and satifying
\begin{equation}
\gamma^5\epsilon^A=-\epsilon^A\,,
\qquad\quad
\gamma^5\epsilon_A=\epsilon_A\,.
\end{equation}
In the chosen basis we also have
\begin{align}
C\:\left(\bar{\epsilon}_{A}\right)^{T}   =\epsilon_{A}
\quad\;\Leftrightarrow\;\quad
C\left(\left(\epsilon^{A}\right)^{\dagger}\gamma^{0}\right)^{T}=\epsilon_{A}
\quad\;\Leftrightarrow\;\quad
\left(\epsilon^{A}\right)^{\ast} =\epsilon_{A}\:.\qquad
\end{align}
\smallskip

\paragraph{SUSY variations.}
The fermionic variations, in a generic $\N=2$ model with Fayet-Iliopoulos terms $\theta_M$, have the general form \cite{Gallerati:2016oyo,Gallerati:2019mzs}:
\begin{equation}
\begin{split}
\delta\Psi^A_\mu
    &=D_\mu\epsilon^A\:+\:\frac{1}{4}\:T^{+}_{\nu\rho}\;\gamma^{\nu\rho}\,\gamma_\mu\:\varepsilon^{AB}\,\epsilon_B\:+\:\mathbb{S}^{AB}\,\gamma_\mu\,\epsilon_B\;,
\\[2ex]
\delta\lambda^{iA}
   &=-\partial_\mu z^i\,\gamma^\mu\,\epsilon^A \:+\:\frac{1}{2}\;g^{i\bar{\jmath}}\;\bar{f}^\Lambda_{\bar{\jmath}}\;\mathcal{I}_{\Lambda\Sigma}\;F^{\Sigma-}_{\mu\nu}\,\gamma^{\mu\nu}\,\varepsilon^{AB}\,\epsilon_B\:+\:W^{iAB}\,\epsilon_B\;.
\label{eq:susyvars}
\end{split}
\end{equation}
The covariant derivatives are written as
\begin{equation}
D_\mu\epsilon^A=\partial_\mu\epsilon^A+\frac{1}{4}\,{\omega_\mu}^{\!\!ab}\,\gamma_{ab}\,\epsilon^A+\frac{i}{2}\,\left(\sigma^2\right)^{A{\!\!}}_B\;\mathbb{A}_\mu^M\,\theta_M\,\epsilon^B+\frac{i}{2}\,\mathcal{Q}_\mu\,\epsilon^A\:,
\end{equation}
where
\begin{equation}
\mathcal{Q}_\mu=\frac{i}{2}\left(\partial_{\bar{\imath}}\mathcal{K}\,\partial_\mu{\bar{z}}^{\bar{\imath}}-\partial_{i}\mathcal{K}\,\partial_\mu{z}^i\right)\;,
\end{equation}
is the Kahler connection, with $\mathcal{K}$ Kahler potential,
and $\mathbb{A}^M$ the electric and magnetic vector potential.%
\footnote{%
We use the convention of \cite{Anabalon:2020pez,Gallerati:2019mzs} though in a mostly plus signature.
}

The explicit form of the quantities in \eqref{eq:susyvars} is given by \cite{Gallerati:2019mzs}
\begin{equation}
\begin{split}
&F^{\pm}_{\mu\nu}=\frac{1}{2}\left(F_{\mu\nu}\pm i\,{}^{\star{\!}}{F}_{\mu\nu}\right)\:,
\qquad
\mathbb{F}^M_{\mu\nu}=\left(F^\Lambda_{\mu\nu}\,,\:G_{\Lambda\mu\nu}\right)\,,
\qquad
\gamma^{\mu\nu}=\gamma^{[\mu}\gamma^{\nu]}\,,
\\[2ex]
&T_{\mu\nu}=L^\Lambda\;\mathcal{I}_{\Lambda\Sigma}\;F^{\Sigma}_{\mu\nu}
    =\frac{1}{2i}\,L^\Lambda\left(\mathfrak{N}-\overbar{\mathfrak{N}}\right)_{\Lambda\Sigma}\,F^{\Sigma}_{\mu\nu}
    =-\frac{i}{2}\left(M_\Sigma\,F^{\Sigma}_{\mu\nu}-L^\Lambda\,G_{\Lambda\mu\nu}\right)%=
    %\\&\:
    =\frac{i}{2}\,\mathcal{V}^M\;\mathcal{C}_{MN}\;\mathbb{F}^{N}_{\mu\nu}\:,
\\[2.25ex]
%&T^{-}_{\mu\nu}=L^\Lambda\;\mathcal{I}_{\Lambda\Sigma}\;F^{\Sigma-}_{\mu\nu}=\frac{i}{2}\,\mathcal{V}^M\;\mathcal{C}_{MN}\;\mathbb{F}^{N-}_{\mu\nu}\:,\qquad\quad
&T^{+}_{\mu\nu}=\bar{L}^\Lambda\;\mathcal{I}_{\Lambda\Sigma}\;F^{\Sigma+}_{\mu\nu}=-\frac{i}{2}\,\overbar{\mathcal{V}}^M\;\mathcal{C}_{MN}\;\mathbb{F}^{N+}_{\mu\nu}\:,
\qquad
\mathcal{V}^M\=e^{\frac{\mathcal{K}}{2}}\,\Omega^M\=\left(L^\Lambda,\,M_\Lambda\right)\,,
\\[2ex]
%&\T^{-}_{i\,\mu\nu}=-\mathcal{V}^M\;\mathcal{C}_{MN}\;\mathbb{F}^{N-}_{\mu\nu}\:,
%\\[2ex]
&\mathbb{S}^{AB}=
    \frac{i}{2}\,\left(\sigma^2\right)^A{\!\!}_C\;\varepsilon^{BC}\;\mathcal{W}\,,
\qquad
\mathcal{W}=\mathcal{V}^M\,\theta_M\,,
\qquad
\mathcal{U}_i^M=\left(\partial_i+\frac{1}{2}\,\partial_i\mathcal{K}\right)\mathcal{V}^M=\left(f_i^\Lambda,\,h_{i\Lambda}\right)\,,
%\\[2ex]
%&\gamma^{\mu\nu}=
\\[2.5ex]
&W^{i\,AB}=i\,\left(\sigma^2\right)_C{\!\!}^B\;\varepsilon^{CA}\;\theta_M\;g^{i\bar{\jmath}}\;\overbar{\mathcal{U}}^M_{\bar{\jmath}}\,,
\qquad
g_{i\bar{\jmath}}=\partial_i \partial_{\bar{\jmath}}\mathcal{K}\,,
\end{split}
\end{equation}
being $\Omega^M=\left(\mathcal{X}^\Lambda,\,\partial_\Lambda\mathcal{F}\right)$ the holomorphic section of the characteristic bundle defined over the manifold,  $\mathbb{C}_{MN}$ the symplectic invariant matrix and
having also used the properties
\begin{equation}
\overbar{\mathfrak{N}}_{\Lambda\Sigma}\,F^{\Sigma-}=\,G^{-}_\Lambda\;, \quad\qquad
L^\Lambda\,\mathfrak{N}_{\Lambda\Sigma}\,=\,M_\Sigma\;.
\end{equation}
The kinetic matrix \,$\mathfrak{N}=\mathcal{R}+i\,\mathcal{I}$\, can be expressed as \cite{Gaillard:1981rj}
\begin{equation}
\mathfrak{N}_{\Lambda\Sigma}=\partial_{\bar{\Lambda}}\partial_{\bar{\Sigma}}\overbarcal{\mathcal{F}}
    +2\,i\;\frac{\operatorname{Im}\left[\partial_{\Lambda}\partial_{\Gamma}\mathcal{F}\right]\;\operatorname{Im}\left[\partial_{\Sigma}\partial_{\Delta}\mathcal{F}\right]\;L^\Gamma\,L^\Delta}{\operatorname{Im}\left[\partial_{\Delta}\partial_{\Gamma}\mathcal{F}\right]\;L^\Delta\,L^\Gamma}\;,
\end{equation}
with \,$\partial_{\Lambda}=\frac{\partial}{\partial\mathcal{X}^\Lambda}$\,, \,$\partial_{\bar{\Lambda}}=\tfrac{\partial}{\partial{\bar{\mathcal{X}}}^\Lambda}$\,.\par\bigskip
The special geometry of T${}^3$ model, which we will be working in, is characterized by a prepotential
\begin{equation}
\mathcal{F}(\mathcal{X}^\Lambda)\:=-\frac{i}{4}\:\left(\mathcal{X}^0\right)^{\frac{1}{2}}\left(\mathcal{X}^1\right)^{\frac{3}{2}}\,,
\label{eq:prepotentialn}
\end{equation}
and is selected among the class of theories discussed in \cite{Anabalon:2020pez} by choosing $\nu=-2$.

To make contact with the model described in section \ref{sec:model}, we choose the FI terms $\theta_M=(\theta_1,\,\theta_2,\,\theta_3,\,\theta_4)$ to be
\begin{equation} \theta_M=\left(\left(\theta_2/3\right)^{-3}(4\,L)^{-4},\:
\theta_{2}\,,\:
0,\:0\right)\,,
\end{equation}
having set $\alpha=0$ in the general class of models considered in \cite{Anabalon:2020pez}. We will further suitably shift the dilaton and rescale the vector fields as described in the same reference.

\section{Analytic solutions for the fixed fluxes}\label{app:roots}
The explicit relation between the roots $P(x_{0})=0$ in \eqref{px} and the boundary
data $(\psi_{1},\psi_{2})$ is given by
\begin{equation}
\begin{split}
&x_{0,1}^{2}=-W+Z\cos(\theta)\,,
\qquad x_{0,2}^{2}=-W+Z\,\cos\left(\theta+\frac{2\pi}{3}\right),
\qquad
x_{0,3}^{2}=-W+Z\,\cos\left(\theta+\frac{4\pi}{3}\right),
\\[1ex]
&W=\frac{\psi_{1}^{-2}}{6}\left(2\psi_{1}^{2}-2\psi_{2}^{2}+1\right),
\qquad\quad Z=\left(
W^{2}+\psi_{1}^{-2}-\frac{\psi_{1}^{-4}\,\psi_{2}^{2}}{9}+\frac{1}{12}\,\psi
_{1}^{-4}\right)^{1/2}
\\[1.5ex]
&\cos(3\,\theta)  =\frac{W^{3}}{Z^{3}}-\frac{\psi_{1}^{-2}}{Z^{3}}-\frac
{1}{2^{3}\,3^{2}}\frac{\left(  4\psi_{2}^{2}-3\right)  \left(2\psi_{2}^{2}-1\right)  }{\psi_{1}^{6}\,Z^{3}}+\frac{1}{2^{2}\,3^{2}}\frac{16\psi_{2}^{2}-21}{\psi_{1}^{4}\,Z^{3}}\:.
\end{split}
\end{equation}
The $x_{0,1}^{2}$ and $x_{0,3}^{2}$ roots are real positive quantities for certain range of values of $\psi_{1}$ and $\psi_{2}$.
When $x<1$, the $x_{0,1}^{2}$ root yields the large and $x_{0,3}^{2}$ the small soliton, and viceversa for $x>1$. Only in the $x<1$ region these two
configurations coalesce in the same object. Around $\psi_{1}=0$, we find that $x_{0,1}$ is
divergent and $x_{0,3}$ is finite.

\section{Global properties of the $D=11$ background at radial infinity}\label{S7id}
Let us consider the spacetime geometry of our $D=11$ solution at radial infinity and restrict to the submanifold consisting of the product of $S^7$ and the interval $[0,\Delta]$ spanned by $\varphi$.
The corresponding metric in this limit has the general form:
\begin{equation}ds^2_{S^7\times [0,\Delta]}=F_1\,\sum_{I=1}^4\left(d\mu_I^2+\mu_I^2\,
(d\varphi_I-\chi_I\,d\varphi)^2\right)+F_2\,d\varphi^2\,,  \label{ds20}\end{equation}
where $F_1,\,F_2$ are functions and $\chi_I$ are constants derived from the $D=11$ metric in the $x\rightarrow 1$ limit.
This eight-dimensional submanifold has the global structure:
\begin{equation}
(S^7\times [0,\Delta])/\sim\label{gi}
\end{equation}
where $\sim $ is an identification defined as:
\begin{equation}
(p,\varphi=0)\in S^7\times [0,\Delta]\,\,\sim\,\,\, (\mathfrak{M}\cdot p,\varphi=\Delta)\in S^7\times [0,\Delta]\,,
\end{equation}
where $\mathfrak{M}$ is a monodromy acting on $S^7$ as derived below.
 We can describe a point in $S^7$ through the coset representative:
 \begin{equation}
\hat{\mathbb{L}}({\bf X})\equiv \left(\begin{matrix}(1-{\bf X}^T{\bf X})^{\frac{1}{2}} & -{\bf X}^T\cr {\bf X} & (1-{\bf X}{\bf X}^T)^{\frac{1}{2}}\end{matrix}\right)\in \frac{{\rm SO}(8)}{{\rm SO}(7)}\,,
\end{equation}
where ${\bf X}=(X^m)_{m=1,\dots, 7}$ is a 7-vector. To obtain the parametrization in terms of $\mu_I,\,\varphi_I$ we write:
\begin{equation}
{\bf X}^T=\left(\mu_1 \cos (\varphi_1),\mu_1 \sin (\varphi_1),\mu_2 \cos (\varphi_2),\mu_2\sin
   (\varphi_2),\mu_3 \cos (\varphi_3),\mu_3 \sin (\varphi_3),\mu_4 \cos (\varphi_4)\right)\,,
\end{equation}
where $\sum_I\mu_I^2=1$. Next we define the restriction
$${\bf X}_0\equiv \left.{\bf X}\right\vert_{\varphi_i=0}\,,$$
which only depends on $\mu_I$ and introduce the matrices:
\begin{equation}
\mathcal{T}(\varphi_I)\equiv\, e^{-({\bf e}_{23}\,\varphi_1+{\bf e}_{45}\,\varphi_2+{\bf e}_{67}\,\varphi_3-{\bf e}_{18}\,\varphi_4)}\,,
\qquad\;
\mathbb{L}_0(\mu_I)\equiv \hat{\mathbb{L}}\big({\bf X}_0(\mu_I)\big)\,,\qquad
\end{equation}
where the matrices ${\bf e}_{ij}\in \mathfrak{so}(8)$ are defined as follows: \begin{equation}
({\bf e}_{ij})^{k\ell}=\delta^k_i\delta^\ell_j-\delta^k_j\delta^\ell_j\,,\qquad\quad
i,j,\dots =1,\dots,8\,.
\end{equation}
One can show the following relation to hold:
\begin{equation}
\mathbb{L}(\mu_I,\,\varphi_I)\equiv \,\mathcal{T}(\varphi_I)\cdot \mathbb{L}_0(\mu_I)=\hat{\mathbb{L}}({\bf X}(\mu_I,\,\varphi_I))\cdot {\bf h}\,,
\end{equation}
where ${\bf h}$ is a local compensating transformation in ${\rm SO}(7)$. Thus both coset representatives $\mathbb{L}(\mu_I,\,\varphi_I)$ and $\hat{\mathbb{L}}({\bf X}(\mu_I,\,\varphi_I))$ yield the same parametrization of $S^7$ in the coordinates $ (\mu_I,\,\varphi_I)$.
The $S^7$-metric in \eqref{ds20} is obtained by \emph{twisting} the coset representative $\mathbb{L}$ by the ${\rm SO}(8)$-transformation $\mathcal{T}(-\chi_I\,\varphi)$, namely by defining a new coset representative as follows:
\begin{equation}
\mathbb{L}'(\mu_I,\,\varphi_I,\,\varphi)\equiv\mathcal{T}(-\chi_I\,\varphi)\cdot \mathbb{L}(\mu_I,\,\varphi_I)=\mathcal{T}(\varphi_I-\chi_I\,\varphi)\cdot \mathbb{L}_0(\mu_I)\,.
\end{equation}
\sloppy
Clearly locally we can write $\mathbb{L}'(\mu_I,\,\varphi_I,\,\varphi)=\mathbb{L}(\mu'_I,\,\varphi'_I)$ and thus this redefinition does not change the local structure of the seven-dimensional manifold. However, as ${\varphi\rightarrow \varphi+\Delta}$, we have:
\begin{equation}
\mathbb{L}'(\mu_I,\,\varphi_I,\,\varphi+\Delta)=\mathfrak{M}\cdot\mathbb{L}'(\mu_I,\,\varphi_I,\,\varphi)\,,
\end{equation}
where the monodromy matrix reads:
\begin{equation}
\mathfrak{M}\equiv\mathcal{T}(-\chi_I\,\Delta)\,.
\end{equation}
This defines the identification \eqref{gi}, as a consequence of which the ${\rm SO}(8)$-symmetry of the sphere is broken to a subgroup $G_0$ which depends on $\chi_I$, being $G_0$ defined as the maximal subgroup of ${\rm SO}(8)$ commuting with $\mathfrak{M}$. For the T${}^3$-model $\chi_2=\chi_3=\chi_4$ and ${\rm SO}(8)$ is broken to ${\rm SO}(2)\times {\rm SO}(6)$.

\section{Remarks on the stability of the hairy soliton vs the AdS soliton at fixed fluxes}
As it can be seen from Figures \ref{fig:fig3}, \ref{fig:fig4}, and \ref{fig:fig5}, the hairy soliton has always larger free
energy at fixed fluxes than the AdS soliton of Horowitz and Myers
\cite{Horowitz:1998ha}. Let us analyze what is the meaning of this for the stability of the hairy solutions.

Besides the radial coordinate, the AdS soliton has two spacelike
coordinates, let us call them $\left(  \varphi_{\textsc{c}},\,\omega\right)$, where
$\varphi_{\textsc{c}}$ is the contractible cycle. When $\varphi_{\textsc{c}}=\varphi$, the AdS soliton
can never have the same boundary conditions than the hairy soliton, since the use of the Stokes theorem is incompatible with trivial electromagnetic fields
$F^{1}=F^{2}=0$. Hence, both solution can coincide only at $\psi_{1}=\psi_{2}=0$, in which case the hairy soliton becomes a domain wall for non-trivial $\phi$.
The domain wall would be unstable when fermions are anti-periodic around $\varphi_{\textsc{c}}$ and stable
for periodic fermions. Indeed, when the magnetic flux vanishes, the domain wall is supersymmetric if and only if the fermions are periodic in $\varphi_{\textsc{c}}$.

The AdS soliton can have the same boundary conditions as the hairy soliton
if $\omega=\varphi\,\ $and $\varphi_{\textsc{c}}\in\lbrack0,\Delta_{z}]$. Then, one needs zero
energy to put two Wilson lines around $\omega$ to match the magnetic fluxes
($\psi_{1},\psi_{2}$) of the hairy soliton. Under these conditions, the hairy
soliton would be stable if periodic boundary conditions are set for the
fermions around $\varphi_{\textsc{c}}$. Indeed, regularity of the fermions around
$\varphi_{\textsc{c}}$ in the AdS soliton require that the fermions are antiperiodic when
they go around this cycle. However in the hairy soliton this cycle is
non-contractible and one is free to set periodic or antiperiodic boundary
conditions there. Hence, the hairy AdS soliton is stable for fixed fluxes if periodic boundary conditions are set on its non-contractible
cycle.

We conclude that for non-trivial magnetic fluxes, the AdS soliton is not the lowest energy configuration of the theory, provided the fermions are antiperiodic in one cycle and periodic in the other spacelike cycle. If fermions are antiperiodic in each spacelike cycle of the boundary, the AdS soliton is the ground state, otherwise hairy solutions can take this role.

In the case where the AdS soliton is the ground state, since antiperiodic boundary conditions are set on both cycles, there are two AdS solitons with boundary
\begin{equation}
ds_\text{bound.}^{2}=-dt^{2}+d\varphi_1^{2}+d\varphi_2^{2}\,,
\end{equation}
which are
\begin{equation}
\begin{split}
ds^{2}&=f_1(r)\,d\varphi_1^{2}+\frac{dr^2}{f_1(r)}+\frac{r^2}{L^2}\left(-dt^{2}+d\varphi_2^{2}\right)\,,
\\[1.5ex]
ds^{2}&=f_2(r)\,d\varphi_2^{2}+\frac{dr^2}{f_2(r)}+\frac{r^2}{L^2}\left(-dt^{2}+d\varphi_1^{2}\right)\,,
\end{split}
\end{equation}
with
\begin{equation}
f_{1,2}(r)=\frac{r^2}{L^2}-\frac{\mu_{1,2}}{r}\:.
\end{equation}These two solutions have the same energy provided $\mu_1=\mu_2$. It is clear that in $p+2$ dimensions there are $p$ solitons with the same energy for boundary conditions where the fermions are antiperiodic along every spacelike cycle.

\newpage

\hypersetup{linkcolor=blue}
\phantomsection % use it for correct TOC link !!!
\addtocontents{toc}{\protect\addvspace{4.5pt}}% add vertical space in TOC
\addcontentsline{toc}{section}{References} % add References to TOC
\bibliographystyle{mybibstyle}
\bibliography{bibliografia}

\end{document}